\pdfoutput=1
\documentclass[12pt,a4paper]{article}
\usepackage{ifthen} 
\newboolean{pdflatex}
\setboolean{pdflatex}{true} 

\newboolean{articletitles}
\setboolean{articletitles}{true} 

\newboolean{uprightparticles}
\setboolean{uprightparticles}{false} 

\newboolean{inbibliography}
\setboolean{inbibliography}{true} 

\usepackage{microtype}
\usepackage{lineno}  
\usepackage{xspace} 
\usepackage{caption} 

\usepackage{graphicx}  
\usepackage{color}
\usepackage{colortbl}
\graphicspath{{./figs/}} 

\usepackage{amsmath} 
\usepackage{amssymb}
\usepackage{amsfonts}
\usepackage{upgreek} 

\newcommand*\patchAmsMathEnvironmentForLineno[1]{%
\expandafter\let\csname old#1\expandafter\endcsname\csname #1\endcsname
\expandafter\let\csname oldend#1\expandafter\endcsname\csname
end#1\endcsname
 \renewenvironment{#1}%
   {\linenomath\csname old#1\endcsname}%
   {\csname oldend#1\endcsname\endlinenomath}%
}
\newcommand*\patchBothAmsMathEnvironmentsForLineno[1]{%
  \patchAmsMathEnvironmentForLineno{#1}%
  \patchAmsMathEnvironmentForLineno{#1*}%
}
\AtBeginDocument{%
\patchBothAmsMathEnvironmentsForLineno{equation}%
\patchBothAmsMathEnvironmentsForLineno{align}%
\patchBothAmsMathEnvironmentsForLineno{flalign}%
\patchBothAmsMathEnvironmentsForLineno{alignat}%
\patchBothAmsMathEnvironmentsForLineno{gather}%
\patchBothAmsMathEnvironmentsForLineno{multline}%
\patchBothAmsMathEnvironmentsForLineno{eqnarray}%
}

\usepackage{hyperref}    
\usepackage[all]{hypcap} 

\usepackage{xspace} 
\usepackage{upgreek}

\def\lhcb {\mbox{LHCb}\xspace}

\def\babar  {\mbox{BaBar}\xspace}

\def\MagUp {\mbox{\em Mag\kern -0.05em Up}\xspace}

\ifthenelse{\boolean{uprightparticles}}%
{

 \def\Ppi         {\ensuremath{\uppi}\xspace}

 \def\PDelta      {\ensuremath{\Delta}\xspace}                 
 \def\PXi      {\ensuremath{\Xi}\xspace}                 
 \def\PLambda      {\ensuremath{\Lambda}\xspace}                 
 \def\PSigma      {\ensuremath{\Sigma}\xspace}                 
 \def\POmega      {\ensuremath{\Omega}\xspace}                 
 \def\PUpsilon      {\ensuremath{\Upsilon}\xspace}                 
 
 \def\PB      {\ensuremath{\mathrm{B}}\xspace}                 
                  
 \def\PD      {\ensuremath{\mathrm{D}}\xspace}

 \def\PK      {\ensuremath{\mathrm{K}}\xspace}

 \def\Pb      {\ensuremath{\mathrm{b}}\xspace}                 
 \def\Pc      {\ensuremath{\mathrm{c}}\xspace}

 \def\Pi      {\ensuremath{\mathrm{i}}\xspace}

 \def\Ps      {\ensuremath{\mathrm{s}}\xspace}

}
{

 \def\Ppi         {\ensuremath{\pi}\xspace}

 \mathchardef\PDelta="7101
 \mathchardef\PXi="7104
 \mathchardef\PLambda="7103
 \mathchardef\PSigma="7106
 \mathchardef\POmega="710A
 \mathchardef\PUpsilon="7107
                  
 \def\PB      {\ensuremath{B}\xspace}                 
                  
 \def\PD      {\ensuremath{D}\xspace}

 \def\PK      {\ensuremath{K}\xspace}

 \def\Pb      {\ensuremath{b}\xspace}                 
 \def\Pc      {\ensuremath{c}\xspace}

 \def\Pi      {\ensuremath{i}\xspace}

 \def\Ps      {\ensuremath{s}\xspace}

}

\makeatletter
\ifcase \@ptsize \relax
  \newcommand{\miniscule}{\@setfontsize\miniscule{4}{5}}
\or
  \newcommand{\miniscule}{\@setfontsize\miniscule{5}{6}}
\or
  \newcommand{\miniscule}{\@setfontsize\miniscule{5}{6}}
\fi
\makeatother

\DeclareRobustCommand{\optbar}[1]{\shortstack{{\miniscule (\rule[.5ex]{1.25em}{.18mm})}
  \\ [-.7ex] $#1$}}

\def\squark    {{\ensuremath{\Ps}}\xspace}

\def\cquark    {{\ensuremath{\Pc}}\xspace}

\def\bquark    {{\ensuremath{\Pb}}\xspace}

\def\pion   {{\ensuremath{\Ppi}}\xspace}
\def\piz    {{\ensuremath{\pion^0}}\xspace}

\def\pip    {{\ensuremath{\pion^+}}\xspace}
\def\pim    {{\ensuremath{\pion^-}}\xspace}

\def\kaon    {{\ensuremath{\PK}}\xspace}
  \def\Kbar    {{\kern 0.2em\overline{\kern -0.2em \PK}{}}\xspace}

\def\KorKbar    {\kern 0.18em\optbar{\kern -0.18em K}{}\xspace}

\def\Kp      {{\ensuremath{\kaon^+}}\xspace}
\def\Km      {{\ensuremath{\kaon^-}}\xspace}

\def\KS      {{\ensuremath{\kaon^0_{\mathrm{ \scriptscriptstyle S}}}}\xspace}

  \def\Dbar    {{\kern 0.2em\overline{\kern -0.2em \PD}{}}\xspace}
\def\D       {{\ensuremath{\PD}}\xspace}

\def\DorDbar    {\kern 0.18em\optbar{\kern -0.18em D}{}\xspace}
\def\Dz      {{\ensuremath{\D^0}}\xspace}
\def\Dzb     {{\ensuremath{\Dbar{}^0}}\xspace}
\def\Dp      {{\ensuremath{\D^+}}\xspace}

\def\Dstar   {{\ensuremath{\D^*}}\xspace}

\def\Dstarz  {{\ensuremath{\D^{*0}}}\xspace}

\def\Dstarp  {{\ensuremath{\D^{*+}}}\xspace}

\def\Ds      {{\ensuremath{\D^+_\squark}}\xspace}

\def\Bbar    {{\ensuremath{\kern 0.18em\overline{\kern -0.18em \PB}{}}}\xspace}

\def\BorBbar    {\kern 0.18em\optbar{\kern -0.18em B}{}\xspace}

  \def\Y#1S{\ensuremath{\PUpsilon{(#1S)}}\xspace}

\def\Lbar        {{\ensuremath{\kern 0.1em\overline{\kern -0.1em\PLambda}}}\xspace}
\def\LorLbar    {\kern 0.18em\optbar{\kern -0.18em \PLambda}{}\xspace}

\newcommand{\decay}[2]{\ensuremath{#1\!\to #2}\xspace}         

\def\to                 {\ensuremath{\rightarrow}\xspace}

\def\AT#1     {\ensuremath{A_{\mathrm{T}}^{#1}}\xspace}           

\def\C#1      {\ensuremath{\mathcal{C}_{#1}}\xspace}                       
\def\Cp#1     {\ensuremath{\mathcal{C}_{#1}^{'}}\xspace}                    
\def\Ceff#1   {\ensuremath{\mathcal{C}_{#1}^{\mathrm{(eff)}}}\xspace}        
\def\Cpeff#1  {\ensuremath{\mathcal{C}_{#1}^{'\mathrm{(eff)}}}\xspace}       
\def\Ope#1    {\ensuremath{\mathcal{O}_{#1}}\xspace}                       
\def\Opep#1   {\ensuremath{\mathcal{O}_{#1}^{'}}\xspace}                    

\newcommand{\tev}{\ensuremath{\mathrm{\,Te\kern -0.1em V}}\xspace}
\newcommand{\gev}{\ensuremath{\mathrm{\,Ge\kern -0.1em V}}\xspace}
\newcommand{\mev}{\ensuremath{\mathrm{\,Me\kern -0.1em V}}\xspace}
\newcommand{\kev}{\ensuremath{\mathrm{\,ke\kern -0.1em V}}\xspace}
\newcommand{\ev}{\ensuremath{\mathrm{\,e\kern -0.1em V}}\xspace}
\newcommand{\gevc}{\ensuremath{{\mathrm{\,Ge\kern -0.1em V\!/}c}}\xspace}
\newcommand{\mevc}{\ensuremath{{\mathrm{\,Me\kern -0.1em V\!/}c}}\xspace}
\newcommand{\gevcc}{\ensuremath{{\mathrm{\,Ge\kern -0.1em V\!/}c^2}}\xspace}
\newcommand{\gevgevcccc}{\ensuremath{{\mathrm{\,Ge\kern -0.1em V^2\!/}c^4}}\xspace}
\newcommand{\mevcc}{\ensuremath{{\mathrm{\,Me\kern -0.1em V\!/}c^2}}\xspace}

\def\mum  {\ensuremath{{\,\upmu\mathrm{m}}}\xspace}

\def\invfb   {\ensuremath{\mbox{\,fb}^{-1}}\xspace}

\newcommand{\stat}{\ensuremath{\mathrm{\,(stat)}}\xspace}
\newcommand{\syst}{\ensuremath{\mathrm{\,(syst)}}\xspace}

\def\gsim{{~\raise.15em\hbox{$>$}\kern-.85em
          \lower.35em\hbox{$\sim$}~}\xspace}
\def\lsim{{~\raise.15em\hbox{$<$}\kern-.85em
          \lower.35em\hbox{$\sim$}~}\xspace}

\def\ptot       {\mbox{$p$}\xspace}
\def\pt         {\mbox{$p_{\mathrm{ T}}$}\xspace}

\def\evtgen     {\mbox{\textsc{EvtGen}}\xspace}

\def\geant      {\mbox{\textsc{Geant4}}\xspace}

\def\photos     {\mbox{\textsc{Photos}}\xspace}

\def\pythia     {\mbox{\textsc{Pythia}}\xspace}

\def\tell1  {TELL1\xspace}
\def\ukl1   {UKL1\xspace}

\usepackage{cite} 
\usepackage{mciteplus}

\usepackage{color} 
\graphicspath{{./}} 

\def\DsJ     {\ensuremath{\D_{sJ}^+}\xspace}
\def\dstarks   {\ensuremath{\Dstarp \KS}\xspace}
\def\dstarzk   {\ensuremath{\Dstarz \Kp}\xspace}

\def\Dzkpi   {\decay{\Dz}{\Km \pip}}
\def\dstartodpi{\decay{\Dstarp}{\Dz \pip}}
\def\dstarztodpi{\decay{\Dstarz}{\Dz \piz}}

\def\Dzkpia {\decay{\Dz}{\Km \pip \pip \pim}}
\def\Dsjonep {\ensuremath{D_{s1}^*(2860)^+}\xspace}
\def\Dsone {\ensuremath{D_{s1}(2536)^{+}}\xspace}
\def\Dstwo {\ensuremath{D^*_{s2}(2573)^{+}}\xspace}

\def\mthetah     {\mbox{$\theta_{\rm H}$}\xspace}

\def\cthetah     {\mbox{$\cos\theta_{\rm H}$}\xspace}

\def\Dsa {\ensuremath{D^*_{s1}(2700)^+}\xspace}
\def\Dsja {\ensuremath{D^*_{sJ}(2700)^+}\xspace}
\def\Dsb {\ensuremath{D_{s3}^{*}(2860)^+}\xspace}
\def\Dsjb {\ensuremath{D_{sJ}^{*}(2860)^+}\xspace}
\def\Dsc {\ensuremath{D_{sJ}(3040)^+}\xspace}
\def\Dsju {\ensuremath{D_{sJ}(3040)^{+}}\xspace}
\def\calB         {{\ensuremath{\cal B}\xspace}}
\def\calBp         {{\ensuremath{\cal R}\xspace}}
\begin{document}


\renewcommand{\thefootnote}{\fnsymbol{footnote}}
\setcounter{footnote}{1}
\begin{titlepage}
\pagenumbering{roman}

\vspace*{-1.5cm}
\centerline{\large EUROPEAN ORGANIZATION FOR NUCLEAR RESEARCH (CERN)}
\vspace*{1.5cm}
\noindent
\begin{tabular*}{\linewidth}{lc@{\extracolsep{\fill}}r@{\extracolsep{0pt}}}
\ifthenelse{\boolean{pdflatex}}
{\vspace*{-2.7cm}\mbox{\!\!\!\includegraphics[width=.14\textwidth]{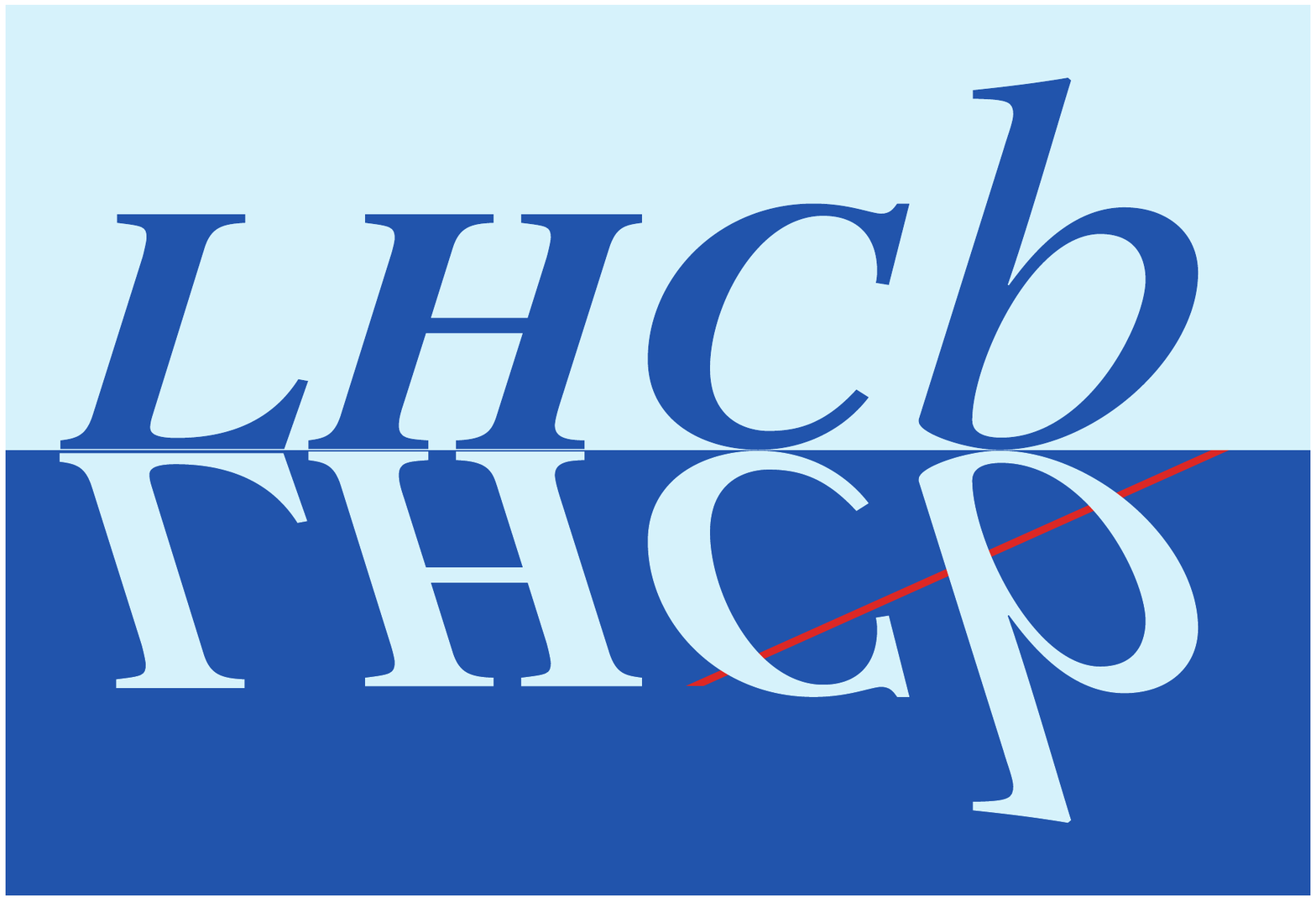}} & &}%
{\vspace*{-1.2cm}\mbox{\!\!\!\includegraphics[width=.12\textwidth]{lhcb-logo.eps}} & &}%
\\
& & CERN-PH-EP-2015-322 \\  
 & & LHCb-PAPER-2015-052 \\  
 & & 26 January 2016 \\
\end{tabular*}

\vspace*{2.0cm}

{\normalfont\bfseries\boldmath\huge
\begin{center}
 Study of $D^{(*)+}_{sJ}$ mesons decaying to $\Dstarp \KS$ and $\Dstarz \Kp$ final states
\end{center}
}

\vspace*{1.0cm}

\begin{center}
The LHCb collaboration\footnote{Authors are listed at the end of this paper.}
\end{center}
\begin{abstract}
  \noindent
  A search is performed for $D^{(*)+}_{sJ}$ mesons in the reactions $pp \to \Dstarp \KS X$ and $pp \to \Dstarz \Kp X$ using data collected at centre-of-mass energies of 7 and $8\,\tev$ with the \lhcb detector.
  For the $\Dstarp \KS$ final state, the decays \dstartodpi with \Dzkpi and \Dzkpia are used. For $\Dstarz \Kp$, the decay \dstarztodpi with \Dzkpi is used.
A prominent \Dsone signal is observed in both $\Dstarp \KS$ and $\Dstarz \Kp$ final states. The resonances \Dsa and \Dsb are also observed, yielding information on their properties, including spin-parity assignments.
The decay \decay{\Dstwo}{\Dstarp \KS} is observed for the first time, at a significance of $6.9\,\sigma$, and its branching fraction relative to the \decay{\Dstwo}{\Dp \KS} decay mode is measured.
\end{abstract}
\vspace*{2.0cm}

\begin{center}
  Submitted to JHEP 
\end{center}

\vspace{\fill}

{\footnotesize 
\centerline{\copyright~CERN on behalf of the \lhcb collaboration, licence \href{http://creativecommons.org/licenses/by/4.0/}{CC-BY-4.0}.}}
\vspace*{2mm}

\end{titlepage}


\newpage
\setcounter{page}{2}
\mbox{~}

\cleardoublepage


\renewcommand{\thefootnote}{\arabic{footnote}}
\setcounter{footnote}{0}

\pagestyle{plain} 
\setcounter{page}{1}
\pagenumbering{arabic}

\section{Introduction}
\label{sec:Introduction}
The discovery by the \babar \ collaboration of a narrow state $D_{s0}^*(2317)^+$ in the decay to $\Ds\piz$~\cite{Aubert:2003fg}, and the subsequent discovery of a second narrow particle, $D_{s1}(2460)^+$ in the decay to $D_s^{*+}\piz$~\cite{Besson:2003cp,Abe:2003jk,Aubert:2003pe}, raised
considerable interest in the spectroscopy of heavy mesons.\footnote{The inclusion of charge-conjugate processes is implied, unless stated otherwise.}
These discoveries
were a surprise because quark model calculations based on heavy quark effective theory (HQET)~\cite{Isgur:1991wq}
predicted the masses of these resonances to be above the $DK$ and $D^*K$ thresholds, respectively.\footnote{In the following $D^*$ is a generic label to indicate the ground state $D^*(2010)^+$ or $D^*(2007)^0$ resonances.}
Consequently their widths
were expected to be very large, as for the corresponding $J^P=0^+$ and $J^P=1^+$ resonances in the $D_J$ spectrum.

The \DsJ mesons are expected to decay into the $D K$ and $D^*K$ final states if they are above threshold.
The BaBar collaboration has explored the $D K$ and $D^*K$ mass spectra~\cite{Aubert:2006mh,Aubert:2009ah}
observing two states, \Dsja and \Dsjb, both decaying to $DK$ and $D^*K$ with a natural parity (NP)
assignment.\footnote{States having $P=(-1)^J$ and therefore $J^P=0^+,1^-,2^+,...$ are referred as
natural parity states and are labelled as $D^*$, while unnatural parity indicates the series $J^P=0^-,1^+,2^-,...$.}
A third structure, \Dsju, is observed only in the $D^*K$ decay mode with a preferred unnatural parity (UP) assignment. The \Dsja resonance was also observed by the Belle and BaBar collaborations in a study of $B$ decays to $\D\Dbar\kaon$~\cite{Brodzicka:2007aa,Lees:2014abp}.
Both collaborations obtain a spin-parity assignment $J^P=1^-$ for this state, and so it is labelled as \Dsa.

The  \lhcb\ experiment has performed studies of the $DK$ final states in the inclusive process, $pp \to DK X$~\cite{Aaij:2012pc}, and in the Dalitz plot analysis of $B_s^0 \rightarrow \Dzb K^- \pi^+$ decays~\cite{Aaij:2014xza,Aaij:2014baa}. 
In the inclusive analysis, the \Dsa and \Dsjb are observed with large statistical significance and
their properties are found to be in agreement with previous measurements. In the exclusive Dalitz plot analysis
of the $B_s^0 \rightarrow \Dzb K^- \pi^+$ decays, the $ \Dzb K^-$ mass 
spectrum shows a complex resonant structure in the $2860\,\mev$ mass region.\footnote{Natural units are used throughout the paper.} This is described by a superposition of a broad $J^P=1^-$ resonance and a narrow $J^P=3^-$ resonance with no evidence for the production of $D_{s1}^*(2700)^-$.
Since the narrow structure at $2860\,\mev$ seen in inclusive $DK$ and $D^*K$ analyses could contain contributions from
various resonances with different spins, it is labelled as \Dsjb.

In references~\cite{Colangelo:2006rq,Close:2006gr,vanBeveren:2006st,Colangelo:2007ds,vanBeveren:2009jq,Colangelo:2012xi} attempts are made to identify these states within the quark model and in Ref.~\cite{Molina:2010tx} within molecular models.
The expected spectrum for $D_s^+$ mesons has recently been recomputed in Refs.~\cite{Godfrey:2013aaa,Godfrey:2015dva}.
In particular, Ref.~\cite{Godfrey:2013aaa} points out that six states are expected in the mass region between 2.7 and $3.0\,\gev$. To date, evidence has been found for three of the states; hence finding the rest would provide an important test of these models.
In this paper we study the \dstarks and \dstarzk systems using $pp$ collision data, collected at centre-of-mass energies of 7 and $8\,\tev$ with the \lhcb detector.

\section{Detector and simulation}
\label{sec:Detector}

The \lhcb detector~\cite{Alves:2008zz,LHCb-DP-2014-002} is a single-arm forward
spectrometer covering the \mbox{pseudorapidity} range $2<\eta <5$,
designed for the study of particles containing \bquark or \cquark
quarks. The detector includes a high-precision tracking system
consisting of a silicon-strip vertex detector surrounding the $pp$
interaction region, a large-area silicon-strip detector located
upstream of a dipole magnet with a bending power of about
$4{\mathrm{\,Tm}}$, and three stations of silicon-strip detectors and straw
drift tubes placed downstream of the magnet.
The tracking system provides a measurement of momentum, \ptot, of charged particles with
a relative uncertainty that varies from 0.5\% at low momentum to 1.0\% at $200\,\gev$.
The minimum distance of a track to a primary vertex (PV), the impact parameter, is measured with a
resolution of $(15+29/\pt)\mum$,
where \pt is the component of the momentum transverse to the beam, in\,\gev.
Different types of charged hadrons are distinguished using information
from two ring-imaging Cherenkov detectors (RICH). 
Photons, electrons and hadrons are identified by a calorimeter system consisting of
scintillating-pad and preshower detectors, an electromagnetic
calorimeter and a hadronic calorimeter. Muons are identified by a
system composed of alternating layers of iron and multiwire
proportional chambers.
The online event selection is performed by a trigger, 
which consists of a hardware stage, based on information from the calorimeter and muon
systems, followed by a software stage, which applies a full event
reconstruction.

In the simulation, $pp$ collisions are generated using
\pythia~\cite{Sjostrand:2006za} 
 with a specific \lhcb
configuration~\cite{LHCb-PROC-2010-056}.  Decays of hadronic particles
are described by \evtgen~\cite{Lange:2001uf}, in which final-state
radiation is generated using \photos~\cite{Golonka:2005pn}. The
interaction of the generated particles with the detector, and its response,
are implemented using the \geant
toolkit~\cite{Allison:2006ve} as described in
Ref.~\cite{LHCb-PROC-2011-006}.
We also make use of simple generator-level simulations~\cite{genbod} to study kinematic effects. 

\section{Event selection}
\label{sec:Selection}

We search for $D^{(*)+}_{sJ}$ mesons using the inclusive reactions
\begin{equation}
pp \to \Dstarp \KS X
\end{equation}
and
\begin{equation}
pp \to \Dstarz \Kp X,
\end{equation}
where $X$ represents a system composed of any collection of charged and neutral particles.
Use is made of both 7 and $8\,\tev$ data for reaction (1), corresponding to an integrated luminosity
of $3\,\invfb$, and $8\,\tev$ data only for reaction (2) which corresponds to an integrated luminosity of $2\,\invfb$.

The charmed mesons in the final state are reconstructed in the decay modes $\Dstarp \to \Dz \pip$, with \Dzkpi and  \Dzkpia, and  $\Dstarz \to \Dz \piz$,
with \Dzkpi and $\piz \to \gamma \gamma$. The \KS mesons are reconstructed in their
$\KS \to \pi^+\pi^-$ decay mode. Because of their long lifetime, \KS mesons may decay inside or outside the vertex detector.
Candidate \KS mesons that are reconstructed using vertex detector information are referred to as ``long'' while those reconstructed without vertex detector information are called ``downstream''.
Those that decay within the vertex detector acceptance have a mass resolution about half as large as those that decay outside of its acceptance.
Reaction (1) with \Dzkpi serves as the primary channel for studying the $D^{(*)+}_{sJ}$ resonance structures and their parameters, while reaction (1) with \Dzkpia and reaction (2) are used for cross-checks and to confirm the observed signatures.

Charged tracks are required to have good track fit quality, momentum $p>3\,\gev$ and $\pt>250\,\mev$. 
These conditions are relaxed to $p>1\,\gev$ and $\pt>150\,\mev$ for the ``soft'' pion originating directly from the \Dstarp decay.
In the reconstruction of the \Dz candidates we remove candidate tracks pointing to a PV, using an impact parameter requirement.
All tracks used to reconstruct the $D$ mesons are required to be consistent with forming a common vertex and the $D$ meson candidate must be consistent with being produced at a PV.
The $\Dstarp$ and $\KS$, and similarly the $D^0$ and $K^+$ candidates, are fitted to a common vertex, for which a good quality fit is required.
The purity of the charmed meson sample is enhanced by requiring the decay products to be identified by the particle identification system, using the difference in the 
log-likelihood between the kaon and pion hypotheses $\Delta\ln\mathcal{L}_{K\pi}$~\cite{Adinolfi:2012an}. 
We impose a tight requirement of $\Delta\ln\mathcal{L}_{K\pi}>3$ for kaon tracks and a loose requirement of $\Delta\ln\mathcal{L}_{K\pi}<10$ for pions. 
The overlap region in the particle identification definition of a kaon and a pion is small and does not affect the measured yields, given the small number of multiple candidates per event. 

Candidate \Dz mesons are required to be within $\pm 2.5\,\sigma$ of the fitted \Dz mass where the mass resolution $\sigma$ is $8.3\,\mev$.
The $\Dz \pip$ invariant mass is computed as
\begin{equation}
m(\Dz \pip) = m(\Km \pip \pip) - m(\Km \pip) + m_{\Dz},
\end{equation}
where $m_{\Dz}$ is the world average value of the \Dz mass~\cite{Agashe:2014kda}. For the channel \Dzkpia, the invariant mass $m(\Dz \pip)$ is defined similarly.

\begin{figure}[ht]
  \begin{center}
    \includegraphics[width=0.45\linewidth]{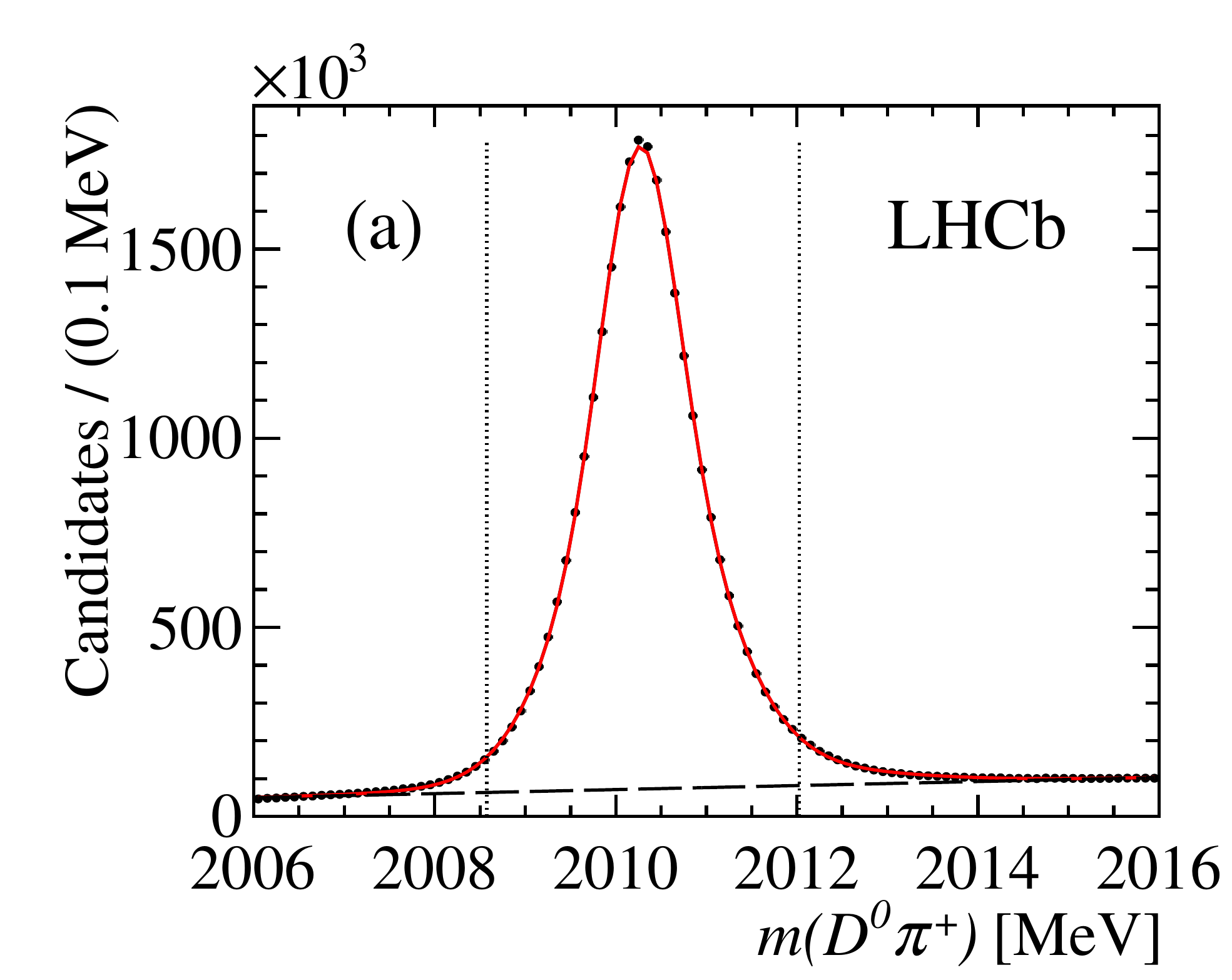}
    \includegraphics[width=0.45\linewidth]{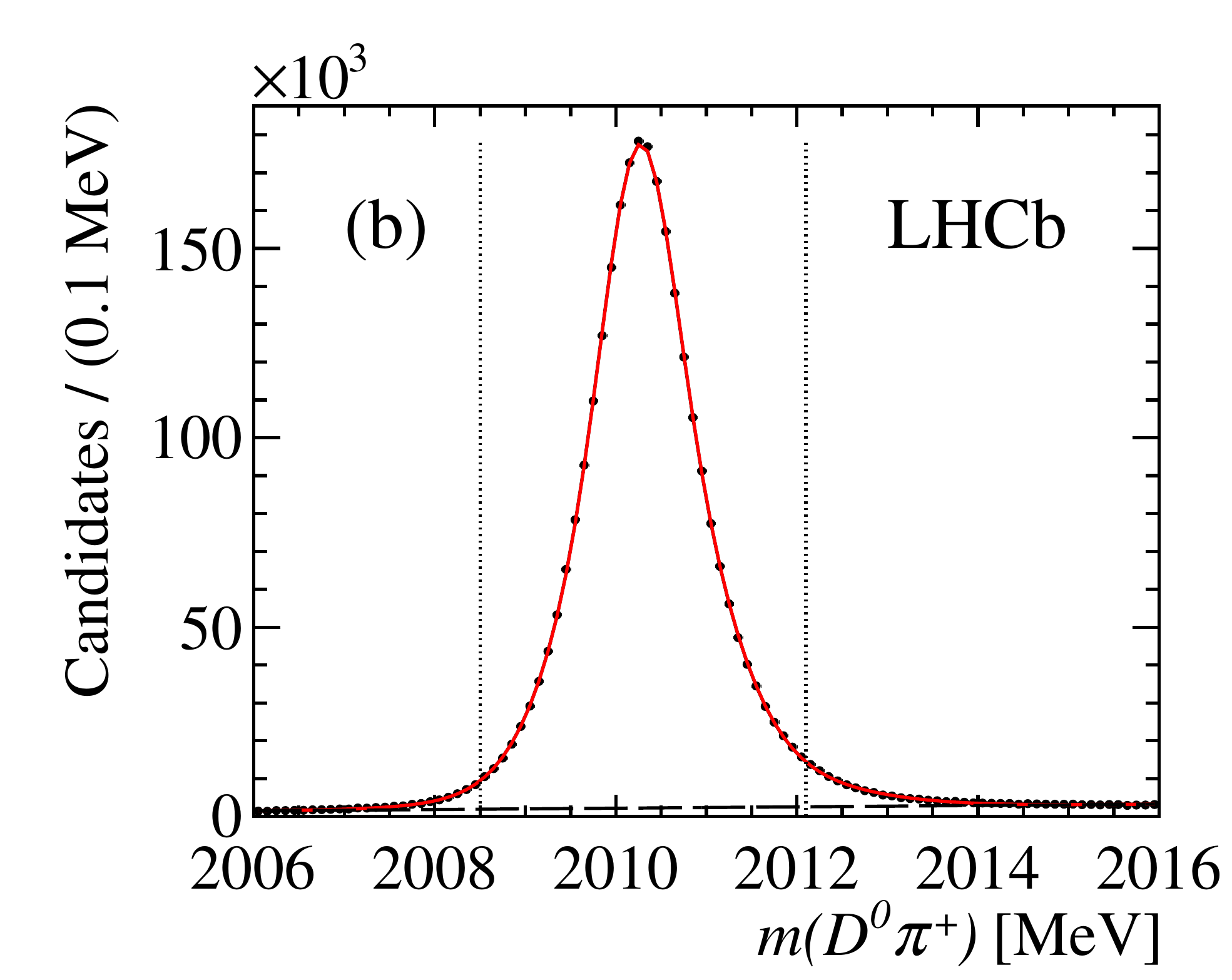}\\
    \includegraphics[width=0.45\linewidth]{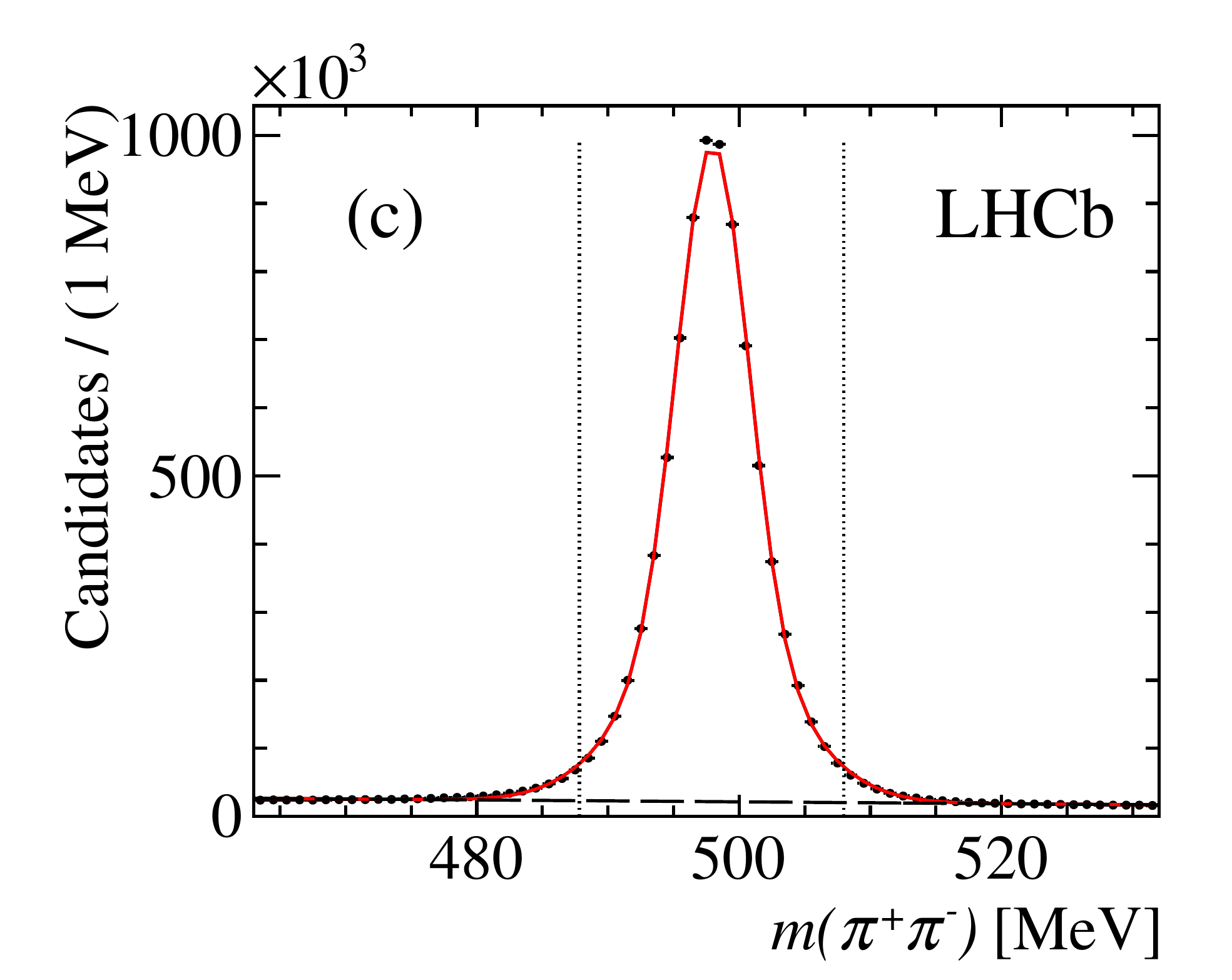}
    \includegraphics[width=0.45\linewidth]{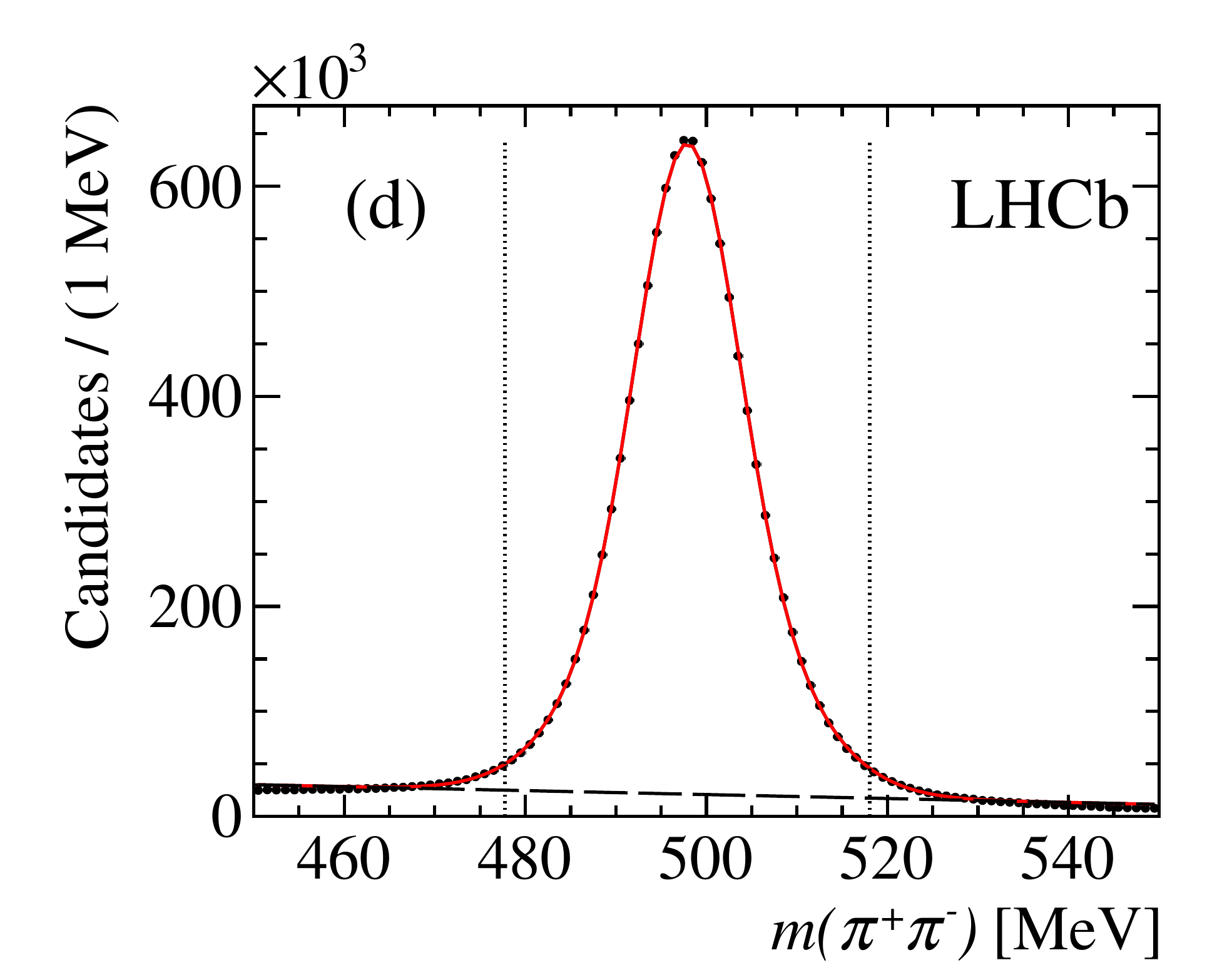}
    \vspace*{-0.5cm}
  \end{center}
  \caption{
    \small Distributions of $\Dz \pip$ invariant mass for (a) \Dzkpi and  (b) \Dzkpia. $\pip \pim$ mass spectrum for (c) long and (d) downstream \KS.
    The full (red) lines describe the fitting function. The dashed lines show the background contributions and the vertical dotted lines indicate the signal regions.
    }
  \label{fig:fig1}
\end{figure}

Figure~\ref{fig:fig1} shows the $\Dz \pip$ invariant mass spectrum for (a) \Dzkpi and  (b) \Dzkpia.
Clean $\Dstarp$ signals for both \Dz decay modes are observed. We fit the mass spectra using the sum of a Gaussian function for the signal and
a second-order polynomial for the background. 
The signal regions are defined to be within $\pm 2.5\,\sigma$ of the peak values,
where $\sigma=0.7\,\mev$ for both channels.

The $\pip \pim$ mass spectra for the two \KS types, the long \KS and downstream \KS, are shown in Figs.~\ref{fig:fig1}(c) and \ref{fig:fig1}(d) and are fitted using the same model as for
the $\Dz \pip$ invariant masses. The signal regions
are similarly defined within $\pm 2.5\,\sigma$ of the peak, with $\sigma=4.1\,\mev$ and $8.7\,\mev$ for long and downstream \KS, respectively.

\begin{figure}[ht]
  \begin{center}
    \hspace*{1.0cm}\includegraphics[width=0.65\linewidth]{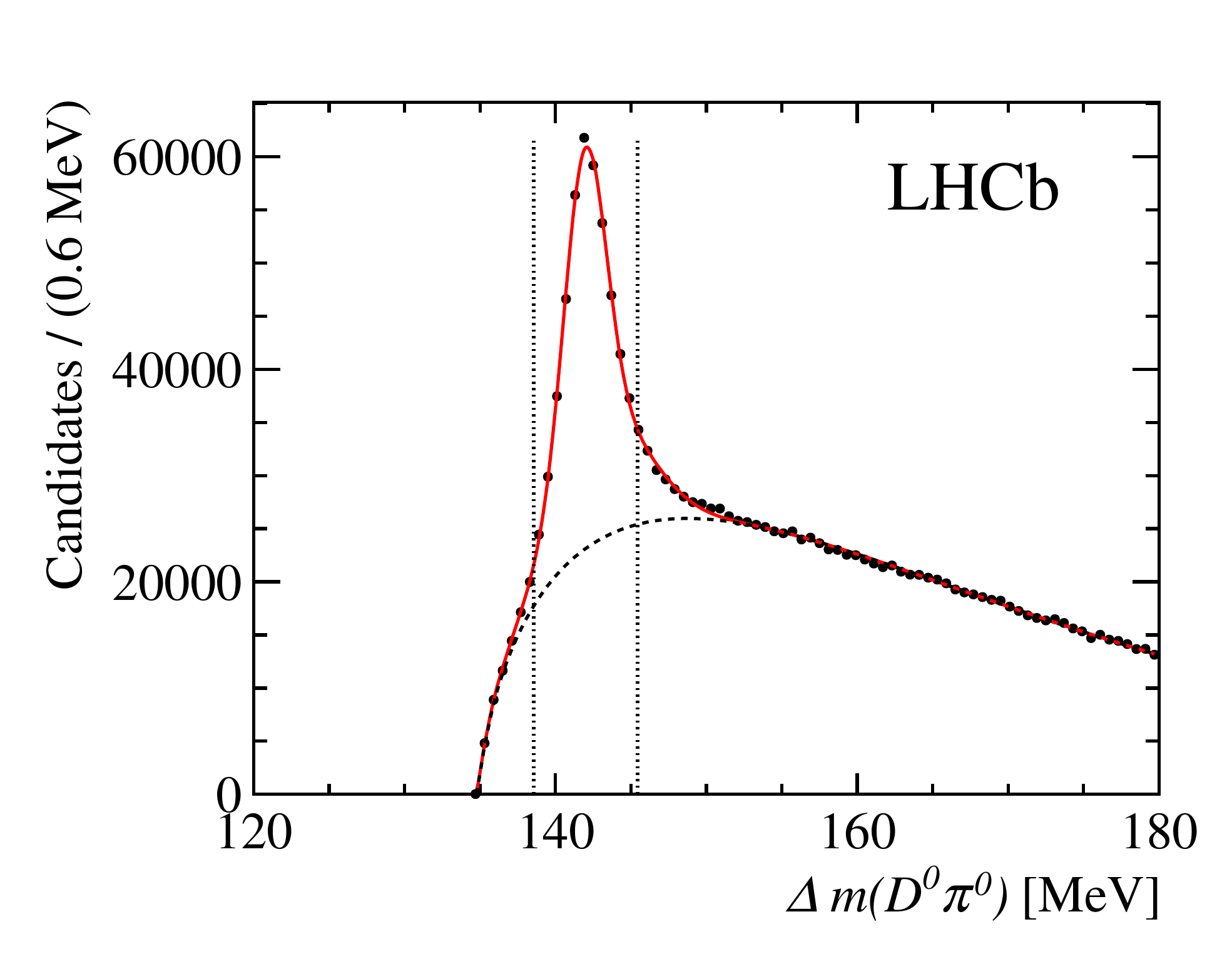}
    \vspace*{-0.5cm}
  \end{center}
  \caption{
    \small Distribution of $\Delta m(\Dz \piz)$ invariant mass. The full (red) line describes the fitting function. The dashed line shows the background contribution and the dotted vertical lines define the $D^{*0}$ signal region.
    }
  \label{fig:fig2}
\end{figure}

The \piz candidates are obtained by kinematically fitting to a $\pi^0$ hypothesis each pair of photon candidates with energy greater
than $600\,\mev$, with the diphoton
mass constrained to the nominal $\pi^0$ mass~\cite{Agashe:2014kda}.
Candidate \Dstarz mesons are formed by combining \Dzkpi decays with all \piz candidates in the event that have $\pt>450\,\mev$.
The resulting \Dstarz candidate is required to have $\pt>6000\,\mev$.
Figure~\ref{fig:fig2} shows the $\Delta m(\Dz \piz) = m(\Km \pip \piz) - m(\Km \pip)$ distribution, where a clear \Dstarz signal can be seen.
The mass spectrum is fitted using for background the threshold function
  \begin{equation}
  B(m) = P(m)(m - m_{th})^\alpha e^{-\beta m - \gamma m^2},
  \label{back2}
  \end{equation}
where in this case $m=\Delta m(\Dz \piz)$, $m_{th}$ is the $\Delta m(\Dz \piz)$ threshold mass and $\alpha$, $\beta$ and $\gamma$ are free parameters. In Eq.~(\ref{back2}) $P(m)$ is the center of mass momentum of the two-body decay of a particle of mass $m$
into two particles with masses $m_1$ and $m_2$,
\begin{equation}
P(m) = \frac{1}{2m} \sqrt{[m^2 - (m_1+m_2)^2][m^2 - (m_1-m_2)^2]}.
\label{eq:ps}
\end{equation}
The function $B(m)$ gives the correct behaviour of the fit at threshold.
The \Dstarz signal is modelled using the sum of two Gaussian functions.
We select the candidates in the $\pm 2\,\sigma$ window around the peak, where $\sigma=1.72\,\mev$ is the width of the dominant Gaussian fitting function, and
we form $D^*K$ pairings by combining \Dstarp and \KS candidates for reaction (1), and \Dstarz and \Kp candidates for reaction (2).

To suppress the large combinatorial background, a set of additional criteria is applied.
We define $\theta$ as the angle between the momentum direction of the kaon in the
$D^*K$ rest frame and the momentum direction of the $D^*K$ system in the laboratory frame. Whereas the signal events are expected to be symmetrically distributed in the variable $\cos\theta$, after correcting for efficiency, more than 90\% of the combinatorial background is found in the negative $\cos\theta$ region.
The  $\cos\theta$ requirements are optimized using the \Dsa signal, an established resonance.
We fit the $D^*K$ mass spectra (using the model described below) with different $\cos\theta$ selections and obtain
the yields for \Dsa signal ($N_{\rm S}$) and background events ($N_{\rm B}$) in the \Dsa signal region (defined in the window $|m(D^*K)-m(\Dsa)|<\Gamma(\Dsa)/2$).
We compute the signal significance $S=N_{\rm S}/\sqrt{N_{\rm S}+N_{\rm B}}$ and signal purity $P=N_{\rm S}/(N_{\rm S}+N_{\rm B})$ and find that the requirements $\cos\theta>0$ (for $\Dstarp \KS$, \Dzkpi), $\cos\theta>-0.15$ (for $\Dstarp \KS$, \Dzkpia) and $\cos\theta>-0.1$ (for $\Dstarz \Kp$) each
provide a good compromise between significance and purity in the respective channel.
With the same method it is also found that it is optimal to require $\pt>4000\,\mev$ for all three final states.
Simulations show that the mass resolution is much smaller than the natural widths of the resonances.

The analysis of the $D^*K$ system, with $D^* \to D \pi$, is a three-body decay and therefore allows a spin analysis of the produced resonances and a 
separation of the different spin-parity components. We define the helicity angle \mthetah as the angle between the $\KS$ and the $\pip$ from the $\Dstarp$ decay, in the rest frame of the $\Dstarp \KS$ system. 
Simulated events are used to determine the efficiency as a function of \cthetah, which is found to be uniform only for the $\Dstarp \KS$ candidates formed from the downstream \KS sample. Therefore, for studying the angular distributions we do not use the long \KS sample, which removes
approximately 30\% of the data. 

\section{Mass spectra}
\label{sec:mass}

\begin{figure}[t]
  \begin{center}
    \hspace*{1.0cm}\includegraphics[width=0.8\linewidth]{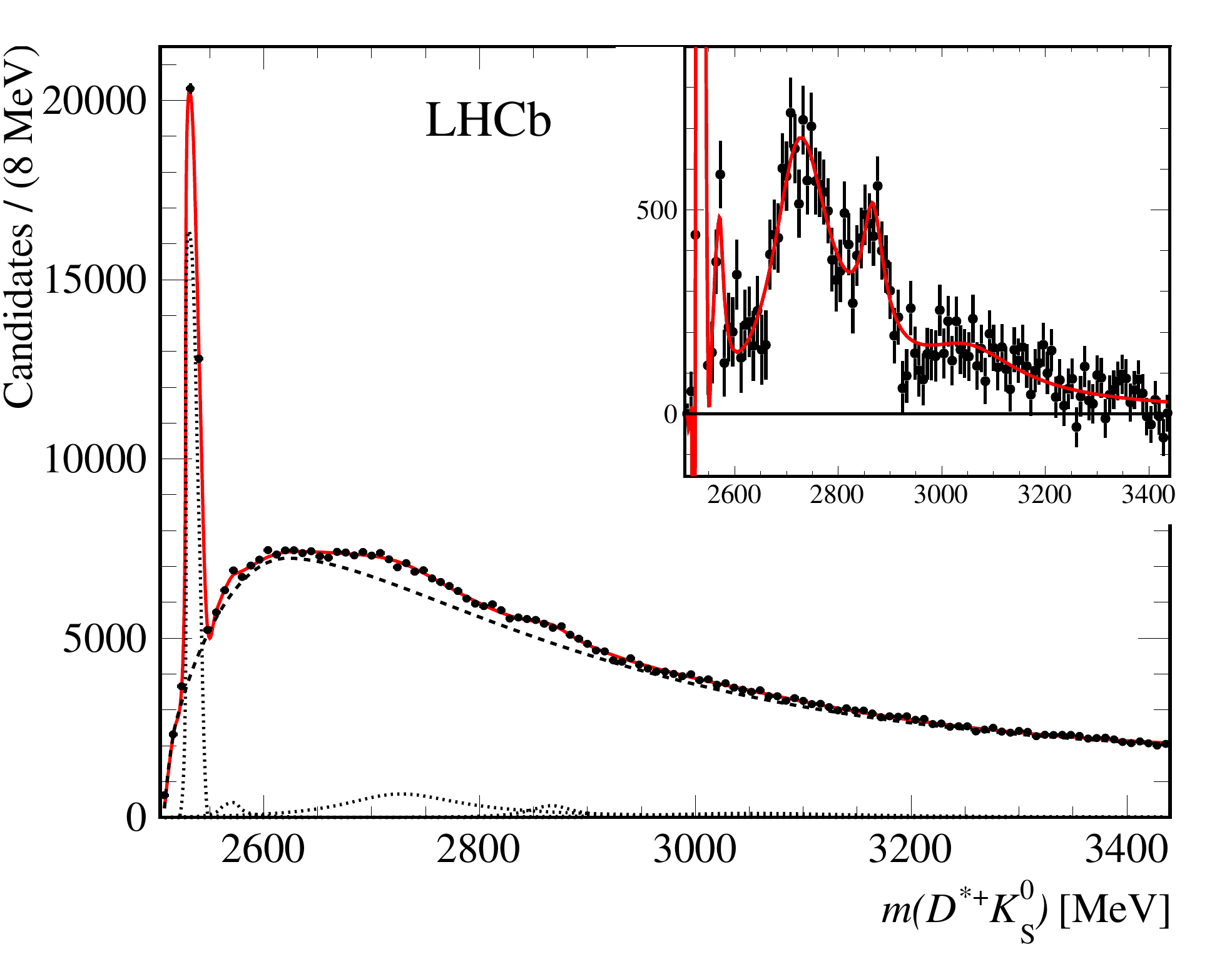}
    \vspace*{-0.5cm}
  \end{center}
  \caption{Distribution of the $\Dstarp \KS$ invariant mass for \Dzkpi decay. The full (red) line describes the fitting function.
  The dashed line displays the fitted background and the dotted lines 
the \Dsone, \Dstwo, \Dsa, \Dsjb and \Dsc contributions. The inset displays the $\Dstarp \KS$ mass spectrum after subtracting the fitted background. 
    \small 
    }
  \label{fig:fig3}
\end{figure}

In order to improve the mass resolution on the $D^*K$ mass spectra, we compute the \Dstarp, \KS and \Dstarz energies using the world average mass measurements~\cite{Agashe:2014kda}.
The $\Dstarp \KS$ mass spectrum for \Dzkpi is shown in Fig.~\ref{fig:fig3} and contains 5.72$\times 10^5$ combinations.
We observe a strong \Dsone signal and weaker resonant contributions due to \Dstwo, \Dsa, and \Dsjb states. The \Dstwo decay to $\Dstarp \KS$ is observed for the first time. 
A binned $\chi^2$ fit to the mass spectrum is performed in which
the narrow \Dsone is described by a Gaussian function with free parameters.
Other resonances are described by relativistic Breit-Wigner~(BW) functions (in $D$-, $P$- and $F$-wave for \Dstwo, \Dsa, and \Dsb respectively).

Using the definition of the center-of-mass momentum $P(m)$ given in Eq.~(\ref{eq:ps}), we parameterize the BW function for a resonance of mass $M$ as
\begin{equation}
BW(m) =  {{P(m) \left({P(m)\over P(M)}\right)^{2L} {D^2(P(M))\over D^2(P(m))} } \over {(m^2-M^2)^2 + M^2\Gamma^2(m)}} ,
\end{equation}
where 
\begin{equation}
\Gamma(m)= \Gamma {M \over m}\left({P(m) \over P(M)}\right)^{2L+1} {D^2(P(M))\over D^2(P(m))},
\end{equation}
and 
\begin{equation}
D(P) = \left\{
\begin{array}{l}
\sqrt{1 +(PR)^2} \ {\rm for} \ L=1,\\
\sqrt{9 + 3(PR)^2 + (PR)^4} \ {\rm for} \ L=2,\\
\sqrt{225 + 45(PR)^2 + 6(PR)^4 + (PR)^6}  \ {\rm for} \ L=3,\\
\end{array}
\right.
\end{equation}
are the Blatt-Weisskopf form factors~\cite{Blatt}.
No dependence of the resonance parameters on the Blatt-Weisskopf radius $R$ is found and it is therefore fixed to $2.5\, \gev^{-1}$.
The quantity $L$ is the angular momentum between the two decay fragments: $L=1$ for $P$-wave, $L=2$ for $D$-wave and $L=3$ for $F$-wave resonances.
The \Dsc resonance is described by a nonrelativistic BW function multiplied by $P(m)$.
The \Dstwo parameters are fixed to the values obtained in the fit to the $D K$ mass spectra~\cite{Aaij:2012pc}.
The background is described by an empirical model~\cite{delAmoSanchez:2010vq}, 
\begin{equation}
B(m) = \left\{
\begin{array}{l}
 P(m)e^{a_1m+a_2m^2} \ {\rm for} \ m<m_0, \\
 P(m)e^{b_0+b_1m+b_2m^2} \ {\rm for} \ m>m_0,
\end{array}
\right.
  \label{eq:back}
\end{equation}
where $P(m)$ is described in Eq.~(\ref{eq:ps}) and $m_0$, $a_{i=1,2}$ and $b_{i=0,1,2}$ are free parameters. In Eq.(\ref{eq:back}) we impose continuity
to $B(m)$ and to its first derivative at the mass $m_0$ and therefore the number of free parameters is reduced by two.
Resonances are included sequentially in order to test the $\chi^2$ improvement when a new contribution is added.
A better fit is obtained if  a broad resonance in the $3000\,\mev$ mass region is included. We find strong correlation between
the parameters of this structure and the background and therefore we add the \Dsc resonance in the fit with parameters fixed
to the values obtained by BaBar~\cite{Aubert:2009ah}.\footnote{$m(\Dsc)=3044 \pm 8\stat^{+30}_{-5}\syst\,\mev$, $\Gamma(\Dsc)=239 \pm 35\stat^{+46}_{-42}\syst\,\mev$.} 

We also study the $\Dstarp \KS$ in the \Dstarp sideband region, defined as $2014.0<m(\Dz \pip)<2018.1\,\mev$. A smooth mass spectrum is obtained, 
well fitted by the above background model with no evidence for additional structures.

Table~\ref{tab:tab1}(a) gives the resulting \Dsa and \Dsjb fitted parameters. 
Statistical significances are computed as $S=\sqrt{\Delta \chi^2}$, where $\Delta \chi^2$ is the difference in $\chi^2$ between fits with 
the resonance included and excluded from the fitting model. Large significances for \Dsa and \Dsjb are obtained, especially
for the \Dzkpi decay mode.
The significance of the \Dsc enhancement is $2.4\,\sigma$.

A search is performed for the \Dsjonep resonance previously observed in the $B_s^0 \to \Dzb \Km \pip$ Dalitz plot analysis~\cite{Aaij:2014xza,Aaij:2014baa}. We first introduce in the fit an incoherent BW function with parameters free to vary within their statistical uncertainties around the reported values in Ref.~\cite{Aaij:2014xza}, but the fit returns a negligible contribution for this state. Since two $J^P=1^-$ overlapping resonances may be
present in the mass spectrum, interference is allowed between the \Dsjonep and the \Dsa resonance by including the amplitude
\begin{equation}
    A_{1^-} = |{\rm BW}_{\Dsa} + c e^{i \phi}{\rm BW}_{\Dsjonep}|^2
    \end{equation}
where $c$ and $\phi$ are free parameters. In this fit we also add the \Dsb resonance with parameters fixed to those from Refs.~\cite{Aaij:2014xza,Aaij:2014baa} and the \Dsa with parameters fixed to those from the $DK$ analysis~\cite{Aaij:2012pc}.
The resulting fit quality is similar to that obtained without the presence of the \Dsjonep resonance ($\chi^2/{\rm ndf}=92/103$). However it is found that the \Dsjonep is accommodated by the fit with strong destructive interference. We conclude
that the data are not sensitive to the \Dsjonep resonance.

\begin{table}
\caption{Results from the fits to the  $\Dstarp \KS$ and $\Dstarz \Kp$ mass spectra. Resonances parameters are expressed in \mev.
When two uncertainties are presented, the first is statistical and the second systematic. The symbol ndf indicates the number of degrees of freedom.}
\label{tab:tab1}
\centering
\resizebox{0.95\textwidth}{!}{
  \begin{tabular}{llccr}
\hline
Data   &  & \Dsa & \Dsjb & $\chi^2/{\rm ndf}$ \cr
\hline
(a) $\Dstarp \KS$  & Mass & $2732.3 \pm 4.3 \pm 5.8$ & $2867.1 \pm  4.3 \pm 1.9$ & \cr
\Dzkpi   & Width & $136 \pm 19 \pm 24$ &  $50 \pm 11 \pm 13$ &  \cr
    & Yield & $(1.57 \pm  0.28) \times 10^4$ & $(3.1 \pm 0.8) \times 10^3$ & 94/103 \cr
& Significance & 8.3 & 6.3 & \cr
\hline
(b) $\Dstarp \KS$  & Mass & $2729.3 \pm 3.3$ & $2861.2 \pm  4.3$ & \cr
\Dzkpi    & Width & 136 (fixed) &  $57 \pm 14 $&  \cr
NP sample & Yield & $(1.50 \pm 0.11) \times 10^4$ & $(2.50 \pm 0.60) \times 10^3$ & 90/104 \cr
& Significance & 7.6 & 7.1 & \cr
\hline
(c) $\Dstarp \KS$  & Mass & 2732.3 (fixed) & $2876.7 \pm  6.4$ & \cr
\Dzkpi    & Width & 136 (fixed) &  $50 \pm 19$ &  \cr
UP sample & Yield & $ (0 \pm 0.8) \times 10^3$ & $(1.0 \pm 0.4) \times 10^3$ & 100/105 \cr
& Significance & 0.0 & 3.6 & \cr
\hline
\hline
(d) $\Dstarp \KS$  & Mass & $2725.5 \pm 6.0$ & $2844.0 \pm  6.5 $ & \cr
\Dzkpia    & Width & 136 (fixed) &  $50  \pm 15 $ &  \cr
    & Yield & $(2.6 \pm  0.4) \times 10^3$ & $490 \pm 180$ & 89/97 \cr
& Significance & 4.7 & 3.8  & \cr
\hline
(e) $\Dstarz \Kp$  & Mass & $2728.3 \pm 6.5$ & $2860.9 \pm  6.0$ & \cr
               & Width & 136 (fixed) &  $50$  (fixed) &  \cr
    & Yield & $(1.89 \pm 0.30) \times 10^3$ & $290 \pm 90$ & 79/99 \cr
& Significance & 6.6 & 3.1 & \cr
\hline
\end{tabular}
}
\end{table}

Systematic uncertainties on the resonance parameters are computed as quadratic sums of the differences between the nominal fit and fits in which the following changes are made.
\begin{itemize}
  \item{} The alternative background function Eq.~(\ref{back2}) is used.
  \item{} The fit bias is evaluated by generating and fitting pseudoexperiments obtained using the parameters from the best fit. The deviations of the mean values of the distributions from the generated ones are taken as systematic uncertainties. 
  \item{} The parameters of the $\Dsc$ state, fixed to the values of Ref.~\cite{Aubert:2009ah} in all the fits, have been varied according
    to their total uncertainties.
  \item{} From the study of high-statistics control samples, a systematic uncertainty of $0.0015\,Q$ on the mass scale is added, where $Q$ is the $Q$-value involved in the resonance decay.
  \item{} The fitting model that includes the \Dsjonep resonance is tested with \Dsb parameters fixed and the \Dsa parameters left free.
\end{itemize}

The different contributions to the systematic uncertainties are summed in quadrature and are summarized in Table~\ref{tab:tab2}. It can be noted that, combining statistical and systematic uncertainties, the resulting \Dsa mass is about $3\,\sigma$ higher than previous measurements while
the \Dsjb parameters are consistent with those of the \Dsb resonance~\cite{Agashe:2014kda}.

  \begin{table}
    \caption{Contributions (in \mev) to the systematic uncertainties on the $\Dsa$ and $\Dsjb$ resonances parameters.}
  \label{tab:tab2}
\centering
\resizebox{0.95\textwidth}{!}{
  \begin{tabular}{lcccc} 
    \hline
    Source &  m($\Dsa$) &    $\Gamma(\Dsa$) &    m($\Dsjb$) &    $\Gamma(\Dsjb$) \cr
    \hline
    Background function & 5.0 & 19.4 & 1.7 & 12.7 \cr
    Fit bias & 0.2 & 1.6 & 0.2 & 1.5 \cr
    $\Dsc$ parameters & 1.3 & 5.7 & 0.5 & 3.2 \cr
    Mass scale & 0.3 &  & 0.5 &  \cr
    Fit model & 2.6 & 12.0 &  & \cr
    \hline
    Total & 5.8 & 23.6 & 1.9 & 13.2 \cr
    \hline
  \end{tabular}
}
  \end{table}

The angular distributions are expected to be proportional to $\sin^2\mthetah$ for NP resonances and proportional to $1+h\cos^2\mthetah$ for UP resonances, where $h$ is a free parameter. 
The $\Dstar K$ decay is forbidden for a $J^P=0^+$ resonance. 
Therefore the selection of candidates
in different ranges of \cthetah can enhance or suppress different spin-parity contributions. We separate the $\Dstarp \KS$ data into two different categories, the NP sample, obtained with the selection $|\cos \theta_{\rm H}|<0.5$ and the 
UP sample, with the selection $|\cos \theta_{\rm H}|>0.5$. 

\begin{figure}[b]
  \begin{center}
    \includegraphics[width=0.45\linewidth]{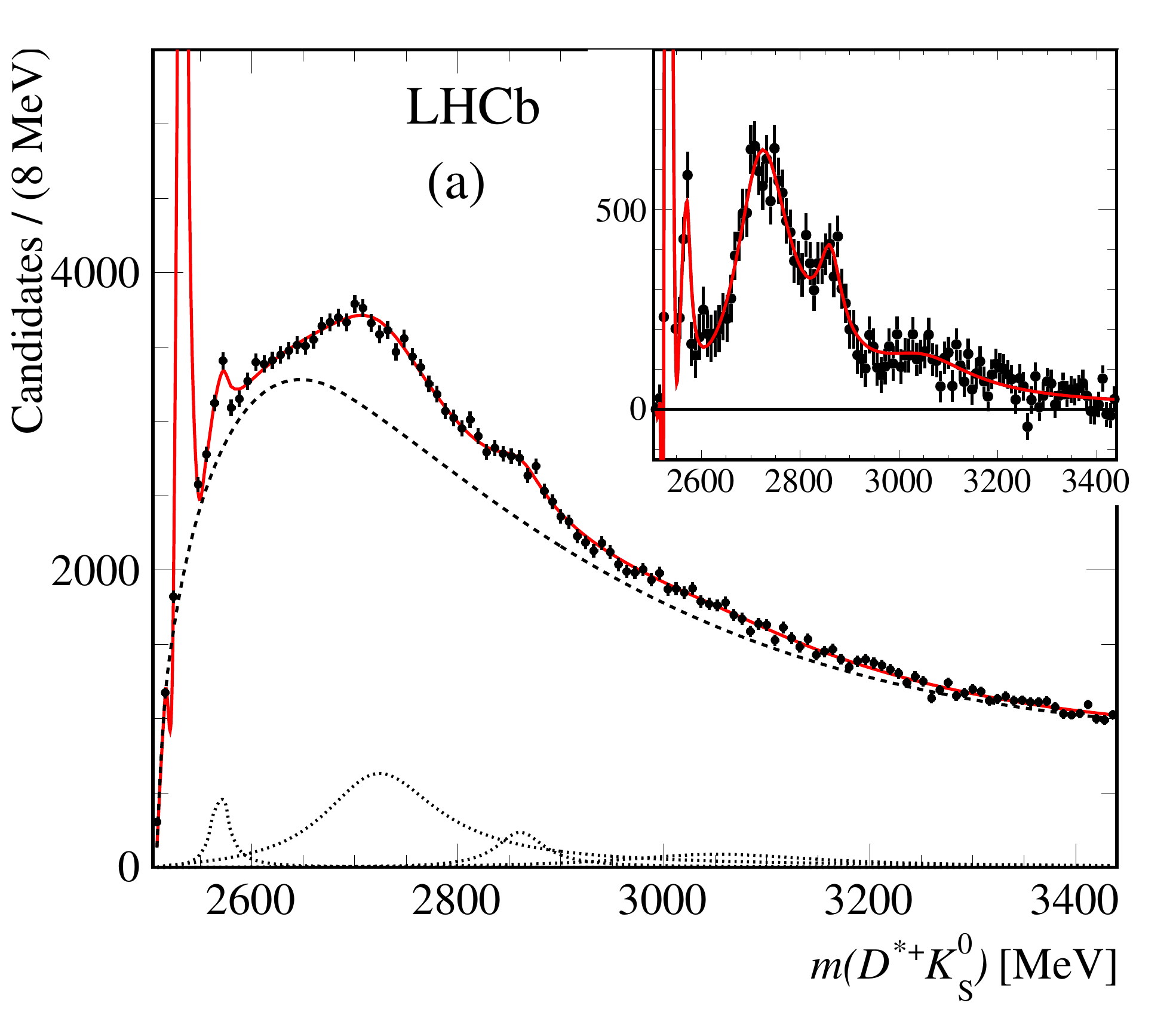}
    \includegraphics[width=0.45\linewidth]{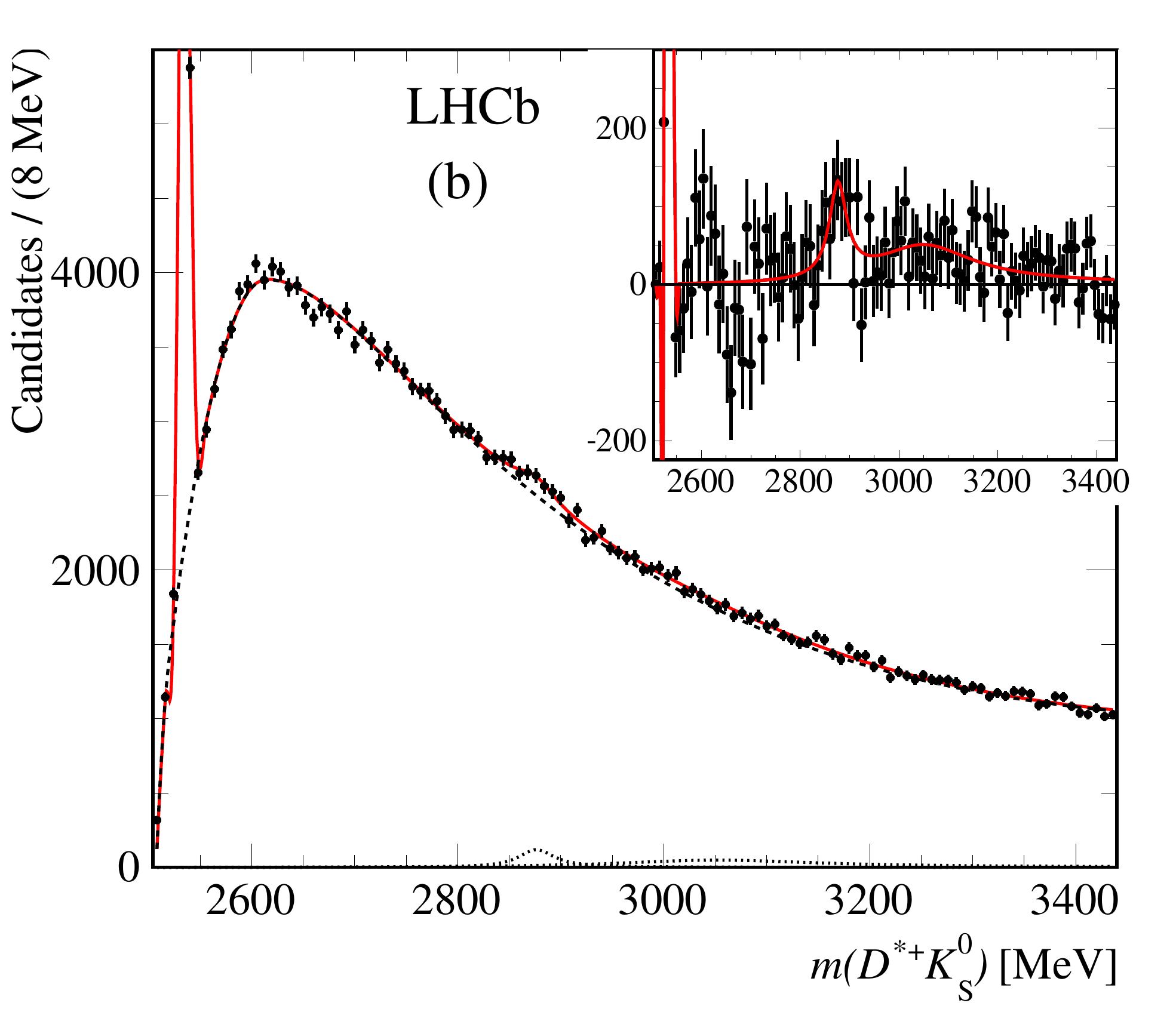}
    \vspace*{-0.5cm}
  \end{center}
  \caption {Mass spectrum of $\Dstarp \KS$ candidates for \Dzkpi in (a) the NP sample, and (b) the UP sample. The full (red) lines describe the fitting function. The dashed lines show the fitted background and the dotted lines 
the \Dstwo, \Dsa, \Dsjb and \Dsc contributions. The insets display the $\Dstarp \KS$ mass spectrum after subtracting the fitted background. 
    }
  \label{fig:fig4}
\end{figure}

The $\Dstarp \KS$ mass spectra for the NP sample is shown in Fig.~\ref{fig:fig4}(a), while the corresponding
mass spectrum for the UP sample is shown in Fig.~\ref{fig:fig4}(b). Most
resonant structures are in the NP sample. An enhancement in the $2860\,\mev$ mass
region in Fig.~\ref{fig:fig4}(b) indicates the possible presence of additional UP contributions.
The fitted parameters are given in Tables~\ref{tab:tab1}(b) and \ref{tab:tab1}(c).

\begin{figure}[t]
  \begin{center}
    \includegraphics[width=0.45\linewidth]{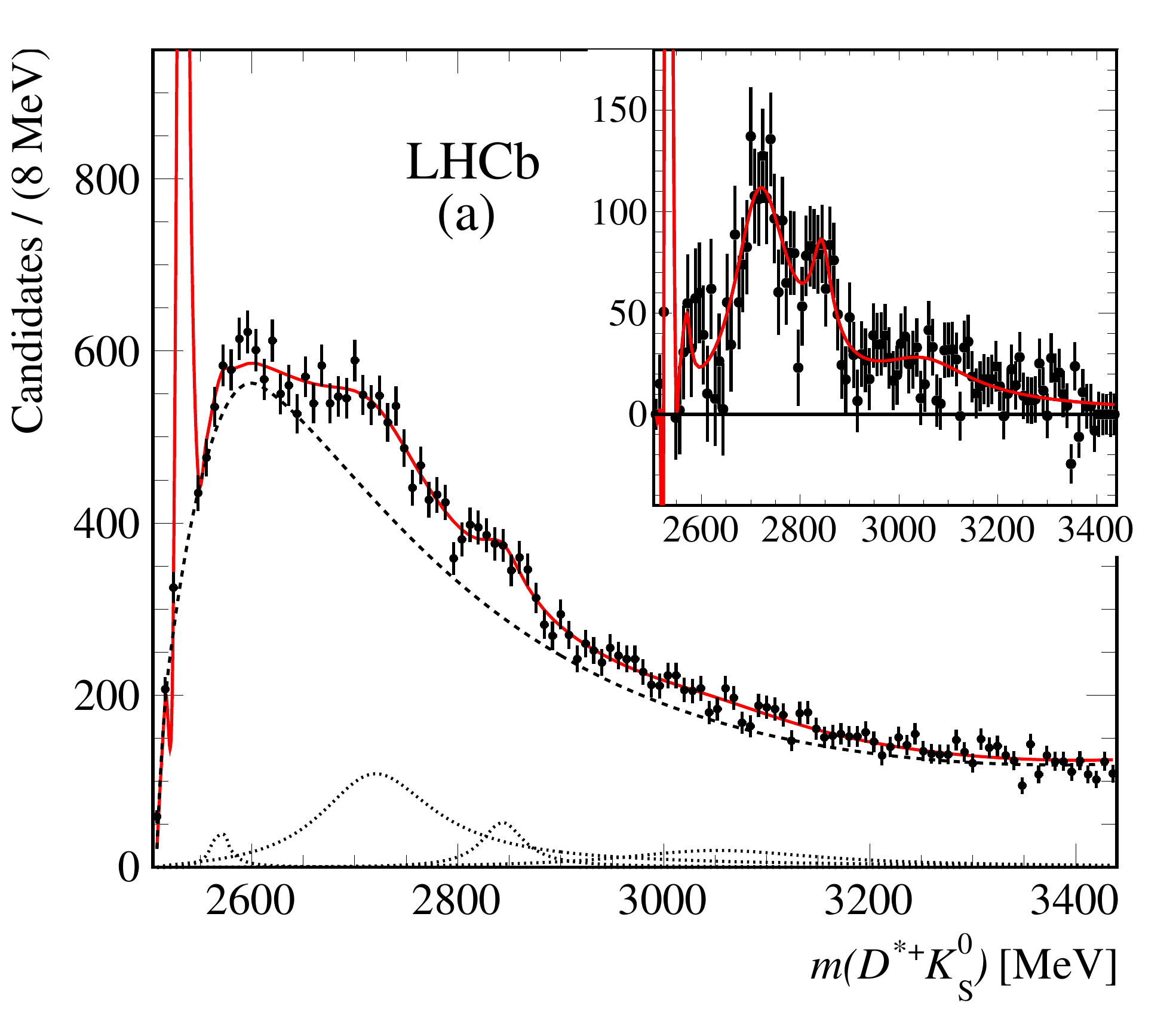}
    \includegraphics[width=0.45\linewidth]{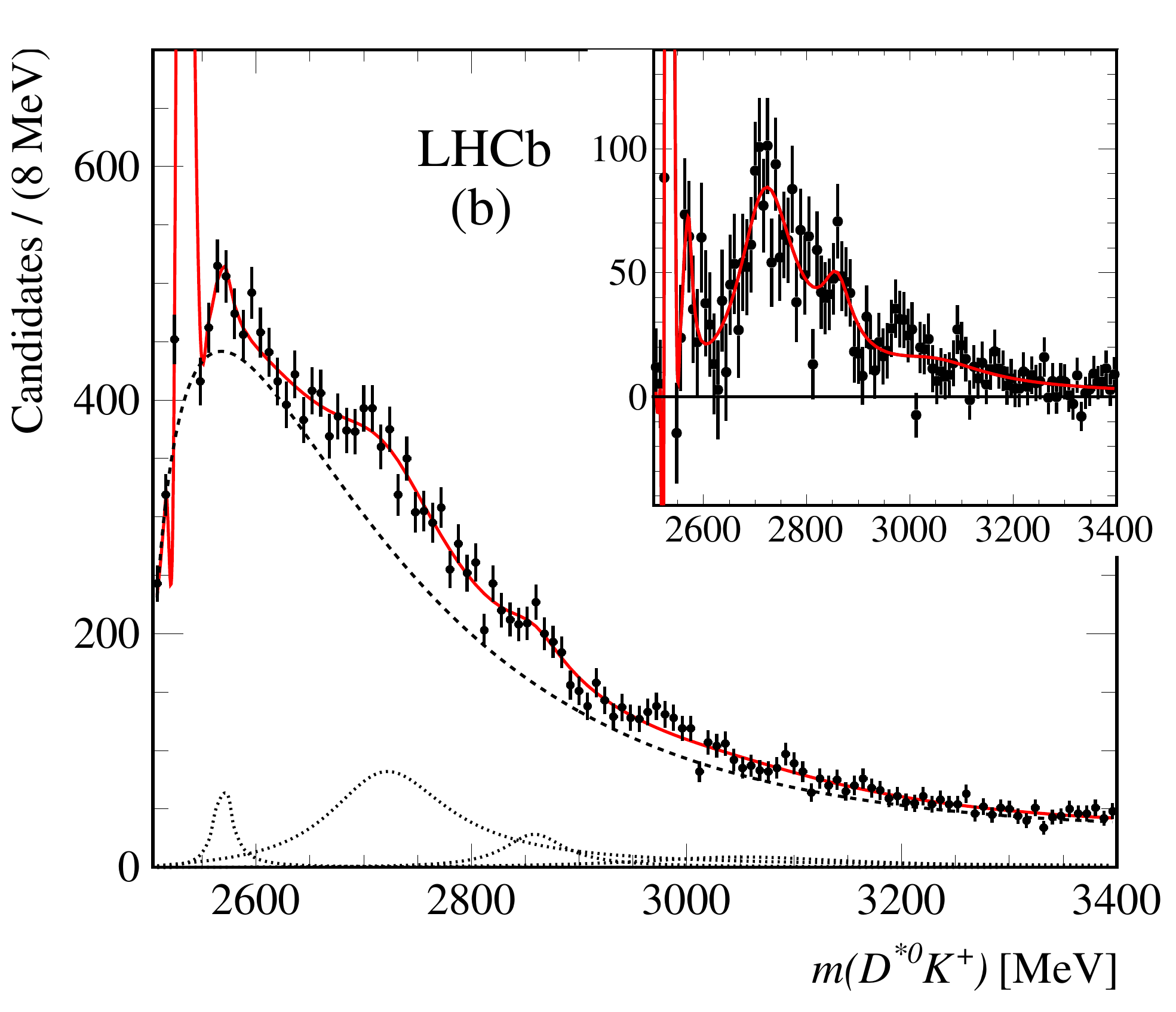}
    \vspace*{-0.5cm}
  \end{center}
  \caption{Mass spectrum of (a) $\Dstarp \KS$ candidates for \Dzkpia, and (b) $\Dstarz \Kp$ candidates for the NP sample.
  The full (red) lines describe the fitting function.
  The dashed lines show the fitted background and the dotted lines 
the \Dstwo, \Dsa, \Dsjb and \Dsc contributions. The insets display the $\Dstarp \KS$ and $\Dstarz \Kp$ mass spectra after subtracting the fitted background. 
    }
  \label{fig:fig5}
\end{figure}

Figure~\ref{fig:fig5}(a) shows the $\Dstarp \KS$ mass spectrum for \Dzkpia, which contains 3.92$\times 10^4$ combinations.
Similar resonant structures to those seen for the $\Dstarp \KS$ final state with \Dzkpi are observed, albeit at lower significance.
Table~\ref{tab:tab1}(d) provides the fitted resonance parameters. Due to the limited data samples, some parameters have been fixed to the values obtained from the fit to the $\Dstarp \KS$ sample with \Dzkpi. The mass values are found to be consistent with the results from the other measurements. 

The $\Dstarz \Kp$ mass spectrum is affected by a high level of combinatorial background, mostly due to the \Dstarz reconstruction (see Fig.~\ref{fig:fig2}).
As observed previously, the $\Dstar K$ mass spectra are dominated by NP resonances and therefore in Fig.~\ref{fig:fig5}(b) we show the $\Dstarz \Kp$ mass spectrum for the NP sample. The mass spectrum contains 2.53$\times 10^4$ combinations.
We observe similar resonant structures as seen in the study of the $\Dstarp \KS$ mass spectra. The fitted resonance parameters are given in Table~\ref{tab:tab1}(e); mass values are consistent with the results from the fits to the other mass spectra.
We do not have the sensitivity to the parameters of the \Dsa and \Dsb resonances in the fits to the $\Dstarp \KS$, \Dzkpia, and $\Dstarz \Kp$ mass spectra
due to the low statistical significance of the signals.

\vspace{0.5cm}
\begin{boldmath}
\section{Measurement of the branching fraction of the decay $\Dstwo \to \Dstarp \KS$}
\label{sec:br}
\end{boldmath}

We measure the branching fraction of the decay $\Dstwo\to \Dstarp \KS$, \Dzkpi relative to that of the decay $\Dstwo \to \Dp \KS$.
For this purpose the $\Dp \KS$
mass spectrum from Ref.~\cite{Aaij:2012pc}, collected at 7~\tev with an integrated luminosity of $1\,\invfb$, is re-fitted.
In this study both long and downstream \KS candidate types are used.
The final states $\Dstarp \KS$, with \Dzkpia and $\Dstarz \Kp$ are used as cross checks and to aid in determining
the significance of the signal.

\begin{figure}[t]
  \begin{center}
    \includegraphics[width=0.80\linewidth]{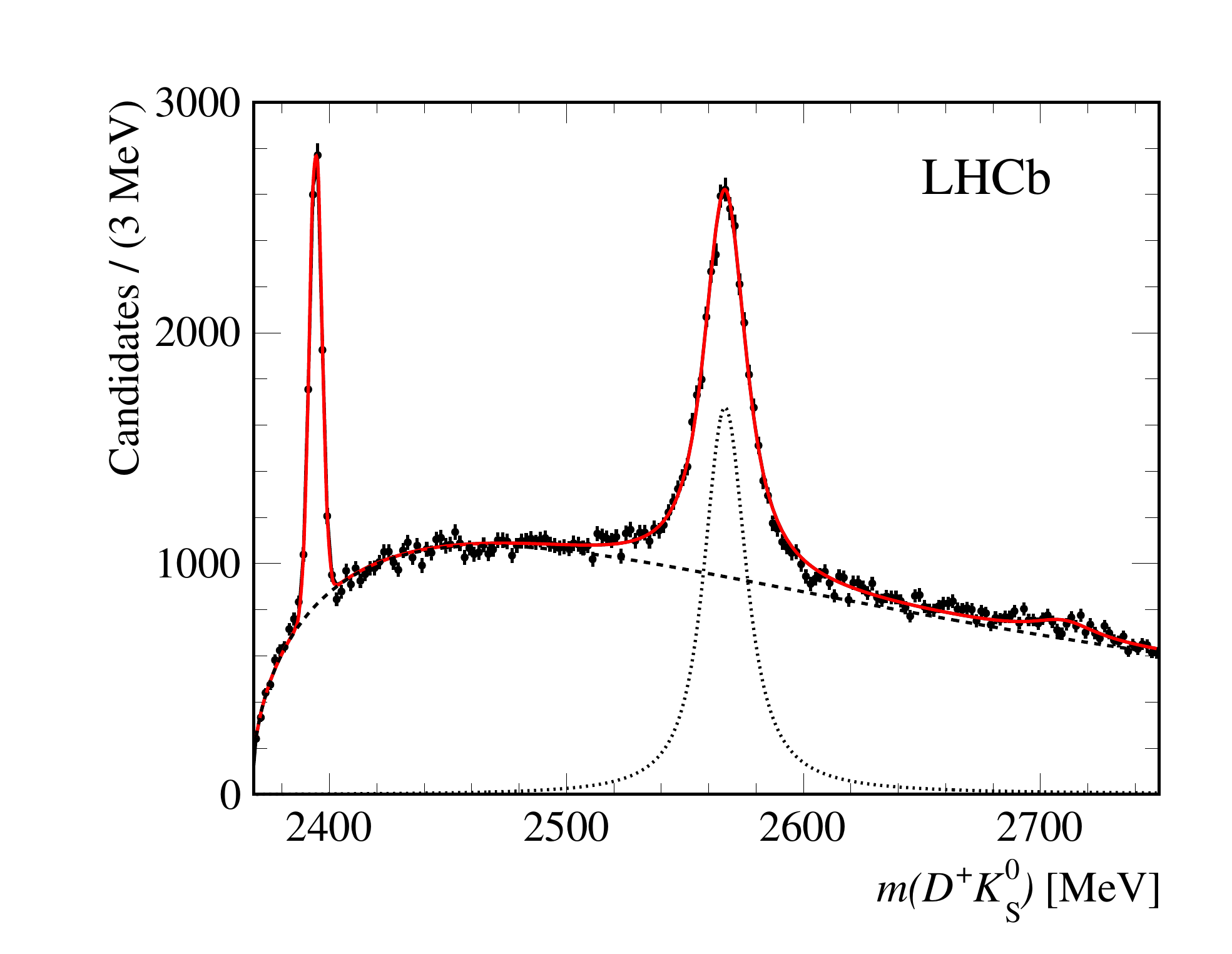}
    \vspace*{-0.5cm}
  \end{center}
  \caption{Distribution of the invariant mass of $\Dp \KS$ candidates from Ref.~\cite{Aaij:2012pc}.
  The full (red) line is the result from the fit described in the text. The dashed line indicates the fitted background and the dotted line shows the fitted \Dstwo contribution.
    }
  \label{fig:fig6}
\end{figure}

Figure~\ref{fig:fig6} shows the $\Dp \KS$ mass spectrum from Ref.~\cite{Aaij:2012pc} along with the results of the fit described below.
A narrow structure is seen near threshold, due to the cross-feed from the decay
\begin{equation}
\Dsone \ \to \KS \Dstarp (\to \Dp \piz/\gamma ),
\end{equation}
where the $\piz/\gamma$ are not reconstructed. In the higher mass region, a strong \Dstwo signal and a weak signal due to the \Dsa resonance are observed.
Due to the difficulty of controlling the systematic uncertainties related to the determination of the relative efficiencies of the $\Dstarp \KS$ and $\Dp \KS$ final states, we normalize the two mass spectra using the \Dsone signal which is observed as a peak in the
$\Dstarp \KS$ and as cross-feed in the $\Dp \KS$ final states.

\begin{figure}[ht]
  \begin{center}
    \includegraphics[width=0.48\linewidth]{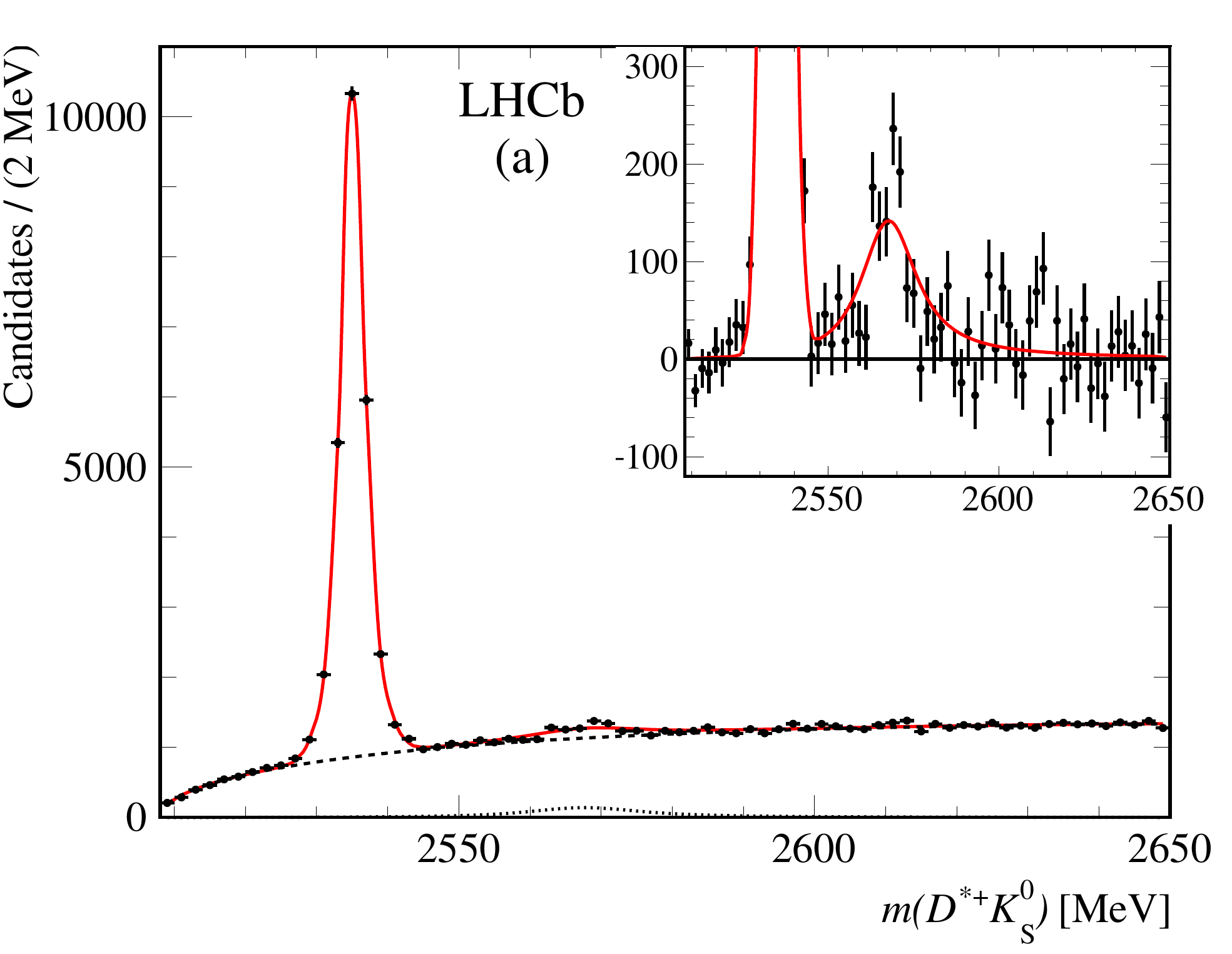}
    \includegraphics[width=0.48\linewidth]{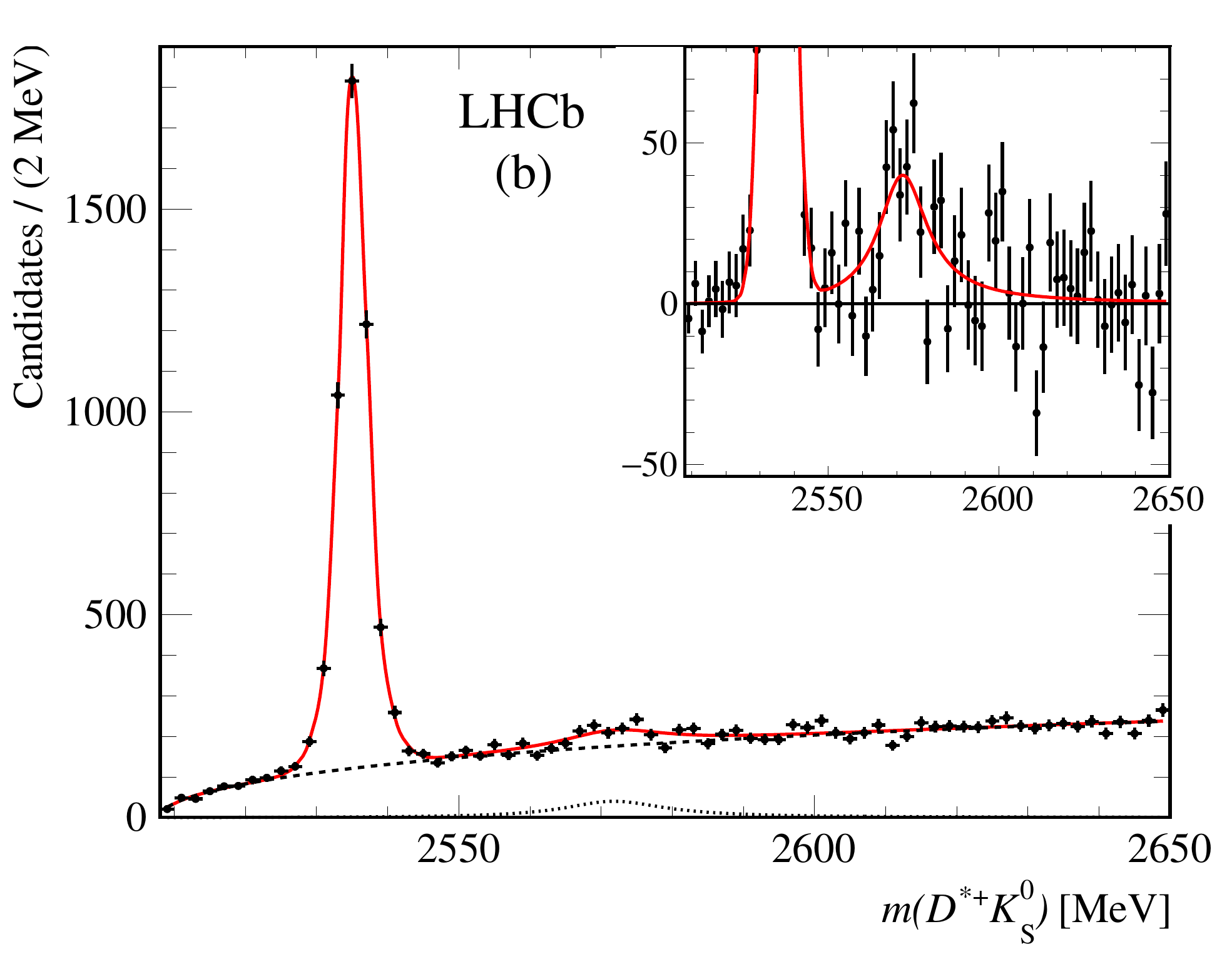}
    \includegraphics[width=0.48\linewidth]{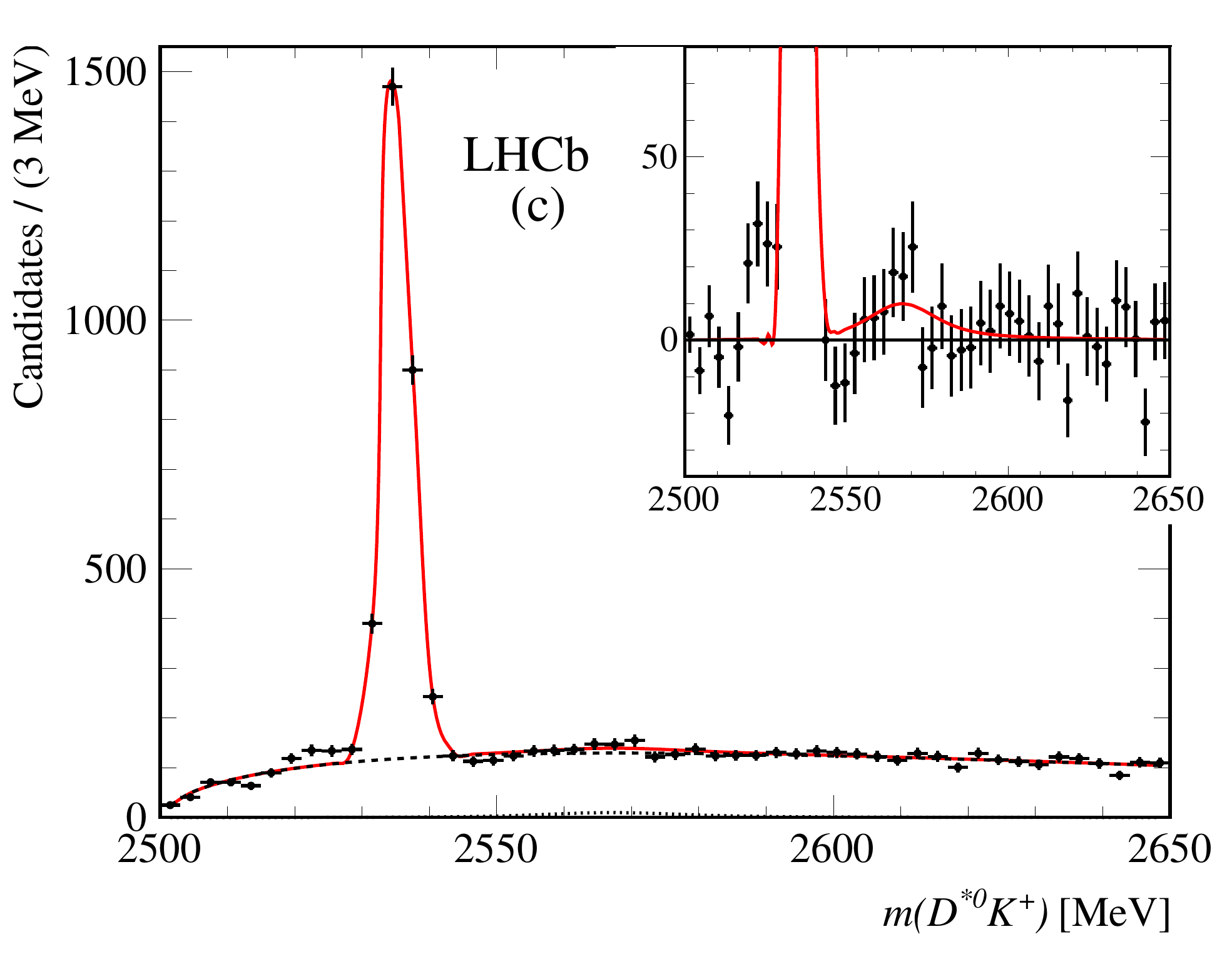}
    \vspace*{-0.5cm}
  \end{center}
  \caption{Mass spectra, in the \Dsone mass region, for the NP sample of (a) $\Dstarp \KS$ with \Dzkpi, (b) $\Dstarp \KS$ with \Dzkpia, and (c) $\Dstarz \Kp$ final states. The full (red) lines describe the fitting function.
  The dashed lines show the fitted background and the dotted lines the \Dstwo contributions. 
The insets display the $\Dstar K$ mass spectra after subtracting the fitted background. 
    }
  \label{fig:fig7}
\end{figure}

The \Dstwo resonance is a well known NP $J^P=2^+$ state. To enhance the signal to
background ratio, we plot in Fig.~\ref{fig:fig7} the $\Dstar K$ mass spectra for the NP sample of the three final states.
All three distributions show a strong \Dsone signal and an enhancement at the \Dstwo mass.

The $\Dp \KS$ mass spectrum and the three $\Dstar K$ mass spectra are fitted using the background function  
\begin{equation}
B(m) = P(m) e^{\beta m + \gamma m^2},
  \label{eq:back1}
\end{equation}
where $P(m)$ is given in Eq.~(\ref{eq:ps}) and $\beta$ and $\gamma$ are free parameters. 
The \Dsone cross-feed into $\Dp \KS$ is modelled using the sum of two Gaussian functions with the same mean, and the
\Dstwo resonance is modelled as a relativistic BW function convolved with a Gaussian function describing the experimental resolution ($\sigma=3.5\,\mev$).
Since the intrinsic width of the \Dsone state in the $\Dstar K$ spectra is much smaller than the experimental resolution,
the \Dsone is modelled using the sum of two Gaussian functions with the same mean.
We obtain $m(\Dsone)=2535.00\pm 0.01\,\mev$, in good agreement with the PDG average.
The \Dstwo resonance is modelled as a relativistic BW function convolved with the experimental resolution ($\sigma=2.5\,\mev$ for \decay{\Dstwo}{\Dstarp \KS}, \Dzkpi)
taking the mass value as a free parameter and with the full width constrained to the value obtained from the fit to the
$\Dp \KS$ mass spectrum ($\Gamma=17.5 \pm 0.4\,\mev$).

 \begin{table}
   \caption{Results from the fits to the $\Dp \KS$ and $\Dstarp \KS$ mass spectra for the evaluation of the $\Dstwo \to \Dstarp \KS$ relative branching fraction.}
  \label{tab:tab3}
\begin{center}
  \begin{tabular}{lccr@{}cc} 
    \hline
    Final state & Mass (MeV) & $\Gamma$ (MeV) &  \multicolumn{2}{c}{Yield} & Significance \cr
    \hline
\decay{\Dstwo}{\Dp \KS} & 2566.9 $\pm$ 0.1 & 17.5 $\pm$ 0.4 \ & (2.55 $\pm$ & \,0.38)$\times 10^4$ &  \cr
 \hline
 \hline
\decay{\Dstwo}{\Dstarp \KS} & 2568.0 $\pm$ 1.0 & 17.5 (fixed) & (2.04 $\pm$ & \,0.26)$\times 10^3$ & $6.9\,\sigma$ \cr
\Dzkpi, NP & & & &  & \cr
 \hline
 \decay{\Dstwo}{\Dstarp \KS} & 2572.0 $\pm$ 1.3 & 17.5 (fixed) & (5.0 $\pm$ & \,1.0)$\times 10^2$ & $4.6\,\sigma$\cr
\Dzkpia, NP & & & & & \cr
\hline
\decay{\Dstwo}{\Dstarz \Kp} & 2567.3 $\pm$ 4.7 & 17.5 (fixed) & (1.1 $\pm$ & \,0.7)$\times 10^2$ & $1.2\,\sigma$\cr
\hline
\hline
\decay{\Dsone}{\Dstarp \KS}  & 2535.00 $\pm$ 0.01 & & (3.59 $\pm$ & \,0.15)$\times 10^4$ & \cr
\Dzkpi, Total & & & & & \cr
 \hline
  \end{tabular}

\end{center}
 \end{table}
 
Table~\ref{tab:tab3} summarizes the fit results.
We note the large statistical significance  of the \Dstwo in the $\Dstarp \KS$ final states, especially for the sample with \Dzkpi.
Consistency is found, within the uncertainties,
in the \Dstwo mass measurements for the different final states. We therefore identify the observed structure as the first observation
of the $\Dstwo \to \Dstarp \KS$ decay.

The relative branching fraction
\begin{equation}
\calBp = \frac{\calB(\Dstwo \to \Dstarp \KS)}{\calB(\Dstwo \to \Dp \KS)}
\end{equation}

\noindent is determined using the results of fits to the  $\Dstwo \to \Dstarp \KS$, $D^0 \to K \pi$ data shown in Fig.~\ref{fig:fig7}(a) and the $\Dstwo \to \Dp \KS$ data shown in Fig.~\ref{fig:fig6}, summarized in Table~\ref{tab:tab3}. 

Using the $\Dstarp \KS$ final state, we verify that the \Dsone cross-feed into the $\Dz \KS$ mass spectrum, when the pion from the \Dstarp decay is ignored, contains all the \Dsone signal. Similarly, using the $\Dstarz \Kp$ data, we ignore the \piz from
the \Dstarz decay and plot the $\Dz \Kp$ mass spectrum. Also in this case, it is found that the \Dsone cross-feed contains all the decays in the \Dsone signal region.
It is assumed that the \Dsone meson decay to $\Dstar K$ is dominant. We test this hypothesis by studying the $\Dz \piz \Kp$ mass spectrum and find that no \Dsone signal is present outside the \decay{\Dstarz}{\Dz \piz} signal region.

Indicating explicitly in brackets the $\Dstarp$ decay modes, we define
\begin{equation}
   R_1 = \frac{N(\Dstwo \to (\Dz \pip) \KS)}{N(\Dsone \to (\Dz \pip) \KS)}
 \end{equation}
and 
 \begin{equation}
   R_2 = \frac{N(\Dsone \to (\Dp \KS)_f)} {N(\Dstwo \to \Dp \KS)},
 \end{equation}
 where $N$ indicates the yields and $\Dsone \to (\Dp \KS)_f$ indicates the cross-feed from $\Dsone \to \Dstarp \KS$ where $\Dstarp \to \Dp (\piz/\gamma)$ and the $\piz/\gamma$ are undetected.

We measure the \Dstwo relative branching ratio as

\begin{equation}
\calBp = R_1 \frac{\epsilon(\Dsone \to (\Dz \pip) \KS)}{\epsilon(\Dstwo \to (\Dz \pip)\KS)}R_2\frac{\epsilon(\Dstwo \to \Dp \KS)}{\epsilon(\Dsone \to (\Dp \KS)_f)} B_Df_{\rm NP},
\label{eq:br}
\end{equation}

\noindent where $\epsilon$ indicates the efficiency for each final state. 
The ratio $B_D$, defined below, is taken from Ref.~\cite{Agashe:2014kda},
\begin{equation} 
B_D = \frac{\calB( \Dstarp \to \Dz \pip)}{\calB( \Dstarp \to \Dp (\piz/\gamma))}=2.10 \pm 0.05,
\end{equation}

\noindent where $\Dp (\piz/\gamma)$ indicates both $\Dp \piz$ and $\Dp \gamma$ decays and $f_{\rm NP}$ is defined below.

In the evaluation of the \Dstwo relative branching fraction, we make use of the $\Dstarp \KS$ NP sample. This selection is used to improve the signal to background ratio for the \Dstwo resonance in the $\Dstarp \KS$ final state.
We also fit the $\Dstarp \KS$ mass spectrum using the full dataset and we report the \Dsone yield indicated as Total in Table~\ref{tab:tab3}. In Eq.~(\ref{eq:br}) the total \Dsone yield is used because of the unnatural parity of this state, and this requires a correction to the \Dstwo yield for the effects
of the NP sample selection. The angular distribution for a NP resonance
is expected to be proportional to $\sin^2 \theta_H$ and therefore the requirement $|\cos \theta_H|<0.5$ selects 69\% of the candidates.
This correction in Eq.~(\ref{eq:br}) is included through the factor $f_{\rm NP}=1.45$.

In Eq.~(\ref{eq:br}) it can be noted that the efficiencies $\epsilon(\Dstwo \to (\Dz \pip)\KS)$ and $\epsilon(\Dsone \to (\Dz \pip) \KS)$
involve the same final state. They are determined from simulation and are found to be the same within uncertainties. 
Similarly, the efficiencies $\epsilon(\Dsone \to (\Dp \KS)_f)$ and
$\epsilon(\Dstwo \to (\Dp \KS))$ are also found to be the same within uncertainties. Therefore, the efficiency ratios are set to unity.

 \begin{table}
   \caption{Measurements used to evaluate the \Dstwo relative branching fraction $\calB(\Dstwo \to \Dstarp \KS)/\calB(\Dstwo \to \Dp \KS)$.}
  \label{tab:tab4}
\begin{center}
  \begin{tabular}{l  r@{}r@{}r} 
    \hline Quantities & \multicolumn{3}{c}{Value} \\
    \hline
    $N(\Dstwo \to (\Dz \pip) \KS)$ & (2.04 $\pm$ & \,0.26\stat $\pm$ & \,0.14\syst) $\times 10^3$\cr
    $N(\Dstwo \to \Dp \KS)$ & (2.55 $\pm$ & \,0.04\stat $\pm$ & \,0.08\syst) $\times 10^4$  \cr
    $N(\Dsone \to (\Dp \KS)_f)$ & (6.54 $\pm$ & \,0.12\stat $\pm$ & \,0.05\syst) $\times 10^3$ \cr
    $N(\Dsone \to (\Dz \pip)\KS) $ & (3.59 $\pm$ & \,0.15\stat $\pm$& \,0.02\syst) $\times 10^4$  \cr
    \hline
    \end{tabular}
    \begin{tabular}{l c}
    $R_1 \qquad \qquad \qquad \qquad \qquad \qquad  $ &  0.057 $\pm$ 0.006\stat $\pm$ 0.004\syst  \cr
    $R_2 \qquad \qquad \qquad \qquad \qquad \qquad $ & 0.256 $\pm$ 0.006\stat $\pm$ 0.008\syst \ \cr
    $f_{\rm NP} \qquad \qquad \qquad \qquad \qquad \qquad $ & 1.45   \cr
    $B_D \qquad \qquad \qquad \qquad \qquad \qquad $ & 2.10 $\pm$ 0.05 \stat   \cr
    \hline
    \end{tabular}
\end{center}
 \end{table}
 
Table~\ref{tab:tab4} summarizes the measurements used to estimate the \Dstwo relative branching fraction.
We obtain
\begin{equation}
\calBp = \frac{\calB(\Dstwo \to \Dstarp \KS)}{\calB(\Dstwo \to \Dp \KS)} = 0.044 \pm 0.005\stat \pm 0.011\syst.
\end{equation}

The systematic uncertainty on the \Dstwo relative branching fraction is computed as the quadratic sum of the differences between the reference values and
those obtained when the following changes are made. 

\begin{itemize}
\item{} The $\Dp \KS$ data are collected at 7~\tev, while the $\Dstar \KS$ data include 7 TeV and 8 TeV data samples. We compute systematic uncertainties
on the $R_1$ and $R_2$ ratios using the $\Dstar \KS$ at 7~\tev only and include the deviation in the systematic uncertainty.
\item{} The uncertainty on the $B_D$ parameter is propagated as a systematic uncertainty.
\item{} Using simulation, we compute efficiency distributions as functions of $m(\Dstarp \KS)$ and $m(\Dp \KS)$ and observe
that they have weak variations in the regions used to evaluate the relative branching fraction.
We assign a 10\% systematic uncertainty to cover the assumptions that the efficiencies 
as functions of $m(\Dstarp \KS)$ and $m(\Dp \KS)$ in Eq.~(\ref{eq:br}) are the same.
\item{} We vary the shape of the background function using Eq.~(\ref{back2}) in the fits to the $\Dstarp \KS$ and $\Dp \KS$ mass spectra and obtain
new estimates for the resonance yields.
We also remove the convolution with the resolution function or replace the relativistic BW functions with
  simple BW functions and include an additional Gaussian function to describe the \Dsone signal.
  \item{} We vary the \Dstwo width by its statistical uncertainty ($0.4\,\mev$) simultaneously in the fits to the $\Dp \KS$ and $\Dstar \KS$ mass spectra. 
\end{itemize}

The contributions to the systematic uncertainty are summarized in Table~\ref{tab:tab5} with 
the dominant component arising from the use of different datasets collected at different centre-of-mass energies.

  \begin{table}
    \caption{Relative systematic uncertainties in the evaluation of the ratio of branching fractions \calBp.}
  \label{tab:tab5}
\begin{center}
  \begin{tabular}{lr} 
    \hline
    Source & Value (\%)\\
    \hline
    Datasets & 22.2 \cr
    Error on $B_D$ & 2.1 \cr
    Efficiency & 10.0 \cr
    Resonance  parameters and backgrounds & 7.5 \cr
    \Dstwo width & 0.3 \cr
    \hline
    Total & 25.6\cr
    \hline
  \end{tabular}
\end{center}
  \end{table}

We also perform a new estimate of the \Dstwo significance in the $\Dstarp \KS$ final state by combining in quadrature the statistical and systematic uncertainties on the yield (see Table~\ref{tab:tab4}) and obtain $S = N_{\rm signal}/\sigma_{\rm tot}=6.9$, where $\sigma_{\rm tot}$ is the total error. This estimate is in good agreement with that reported in Table~\ref{tab:tab3}. 

\vspace{0.5cm}

\begin{boldmath}
\section{Spin-parity analysis of the $\Dstarp \KS$ system}
\end{boldmath}
\label{sec:spin}

We obtain information on the spin-parity of the states observed in the  $\Dstarp \KS$ mass spectrum.
The data for \Dzkpi are first divided into five equally spaced bins in $\cthetah$.
The five mass spectra in the $\Dstarp \KS$ threshold region ($m(\Dstarp \KS)<2650\,\mev$) are fitted using the model described
in Sec.~\ref{sec:br} with fixed \Dsone and \Dstwo resonance parameters, to obtain the signal yields as functions of $\cthetah$ for each resonance.

\begin{figure}[t]
  \begin{center}
    \includegraphics[width=0.49\linewidth]{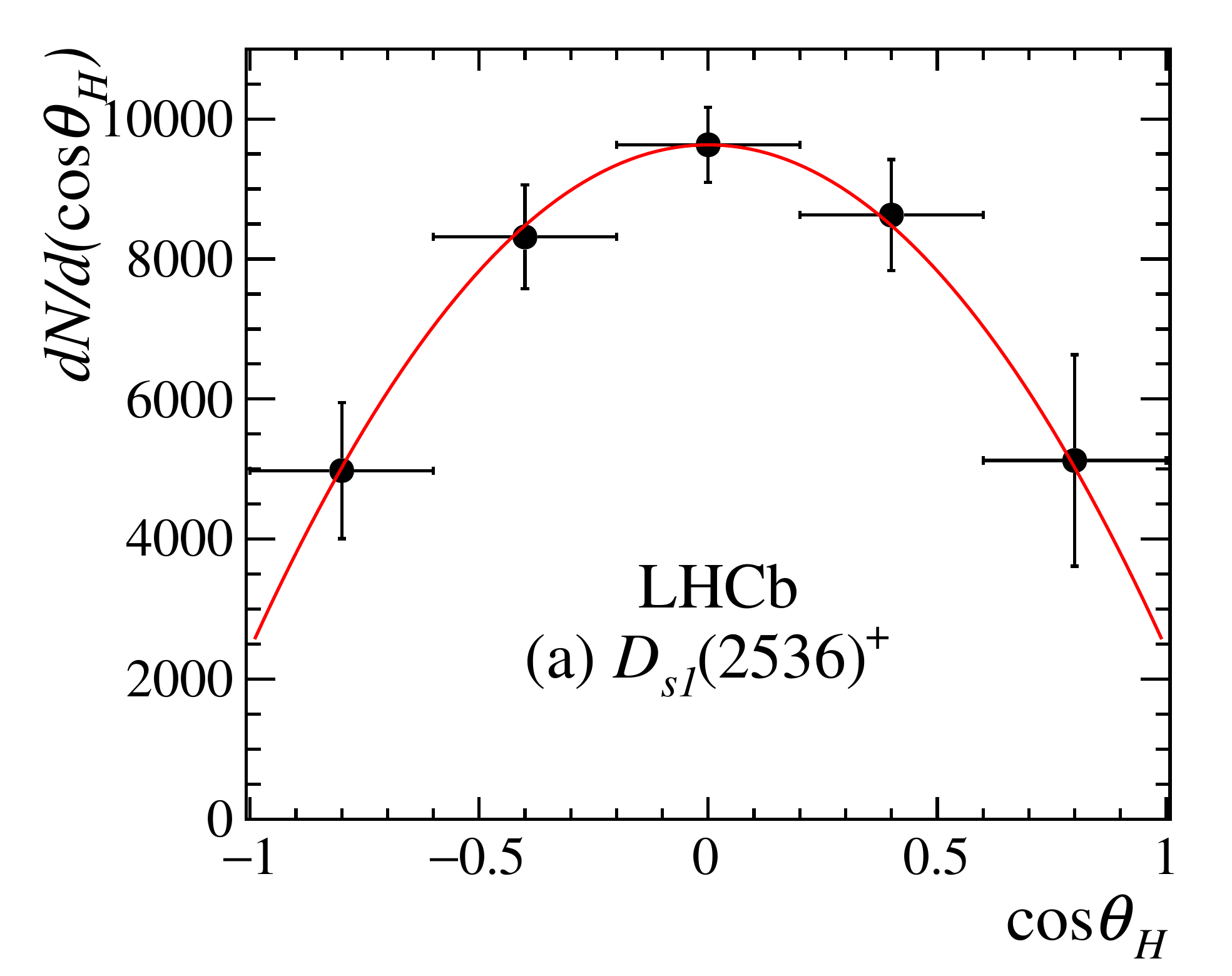}
    \includegraphics[width=0.49\linewidth]{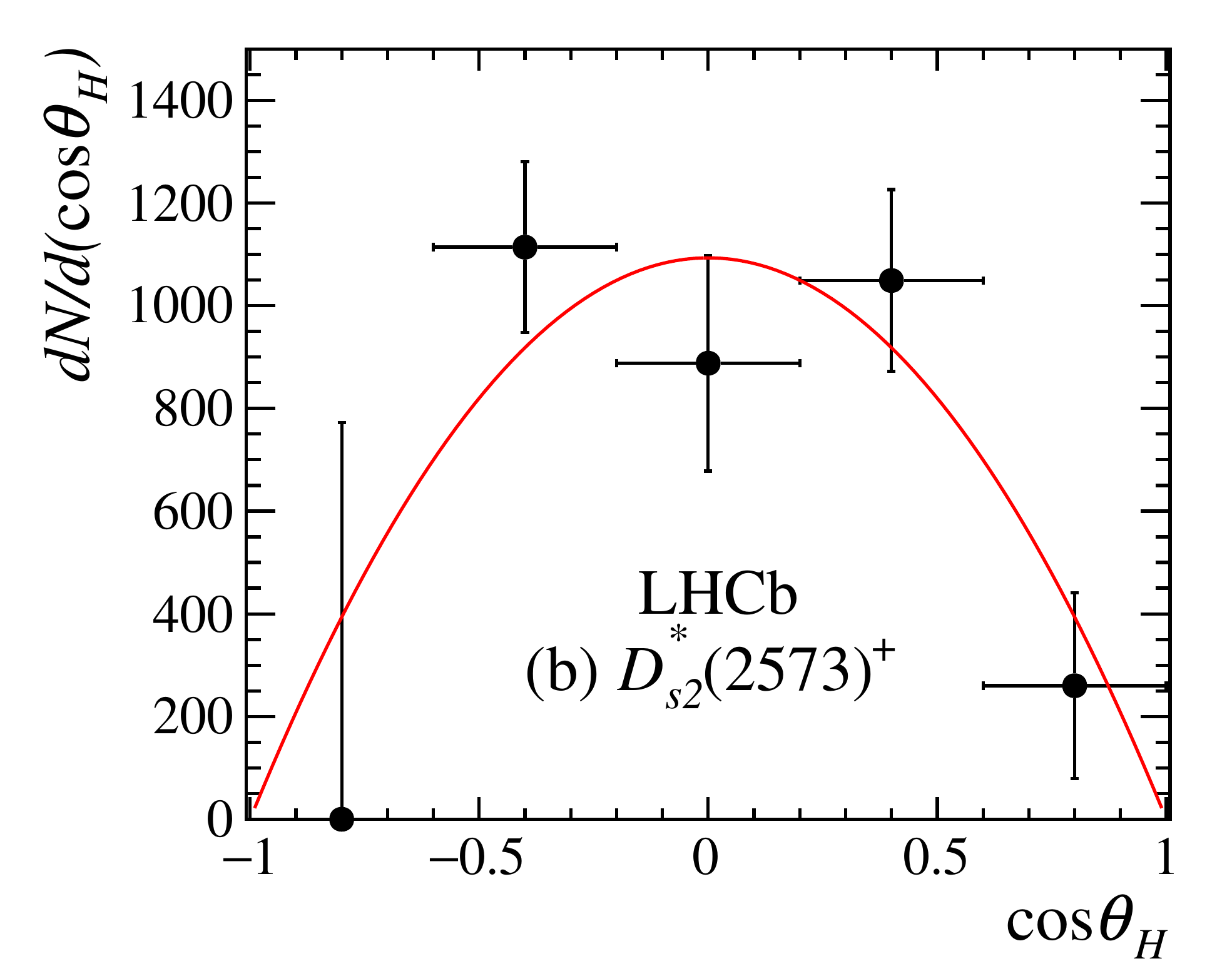}
    \vspace*{-0.5cm}
  \end{center}
  \caption{
      \small Distributions of the measured signal yields for (a) \Dsone and (b) \Dstwo as a function of the 
helicity angle \cthetah. The distributions are fitted with UP (a) and NP (b) functions.
    }
  \label{fig:fig8}
\end{figure}

As stated previously, we determine from simulations that the efficiency as a function of $\cos \theta_H$ is consistent with being uniform; therefore
we plot uncorrected angular distributions.
The resulting distributions for \Dsone and \Dstwo are shown in Fig.~\ref{fig:fig8}(a) and Fig.~\ref{fig:fig8}(b), and
are fitted using the functions
described in Table~\ref{tab:tab6}. A good description of the data is obtained in terms of the expected angular distributions for $J^P=1^+$ and $J^P=2^+$ resonances. We note that the shape of the \Dsone angular distribution is in agreement with that measured in Ref.~\cite{Balagura:2007dya}.

\begin{table}
\caption{Values of $\chi^2/{\rm ndf}$ from the fits to the helicity angles distributions.}
\label{tab:tab6}
\begin{center}
\begin{tabular}{lccr}
\hline
 Resonance  & $J^P$ & Function & $\chi^2/{\rm ndf}$ \cr  
\hline
\Dsone & $1^+$ &  $ 1+h\cos^2\theta_{\rm H}$ & 0.1/3 \cr
\Dstwo & $2^+$ & $\sin^2\theta_{\rm H}$ & 2.2/4  \cr
\Dsa & $1^-$ & $\sin^2\theta_{\rm H}$ & 11.4/7  \cr
\Dsb & $3^-$ & $\sin^2\theta_{\rm H}$ & 13.4/7  \cr
\Dsju & UP & $ 1+h\cos^2\theta_{\rm H}$ & 8.0/6 \cr
\hline
   \end{tabular} 
\end{center}
\end{table}

\begin{figure}[b]
  \begin{center}
    \includegraphics[width=0.45\linewidth]{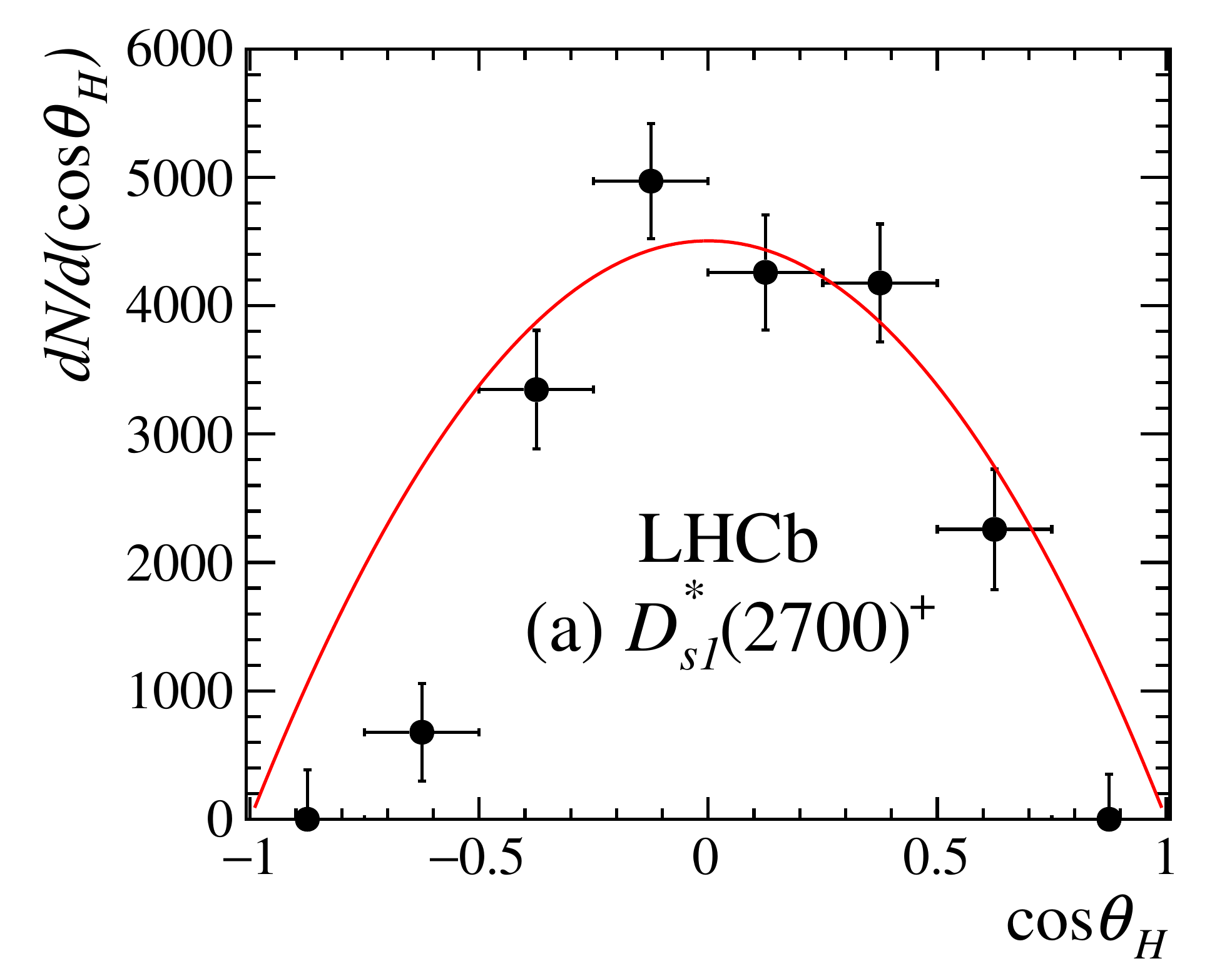}
    \includegraphics[width=0.45\linewidth]{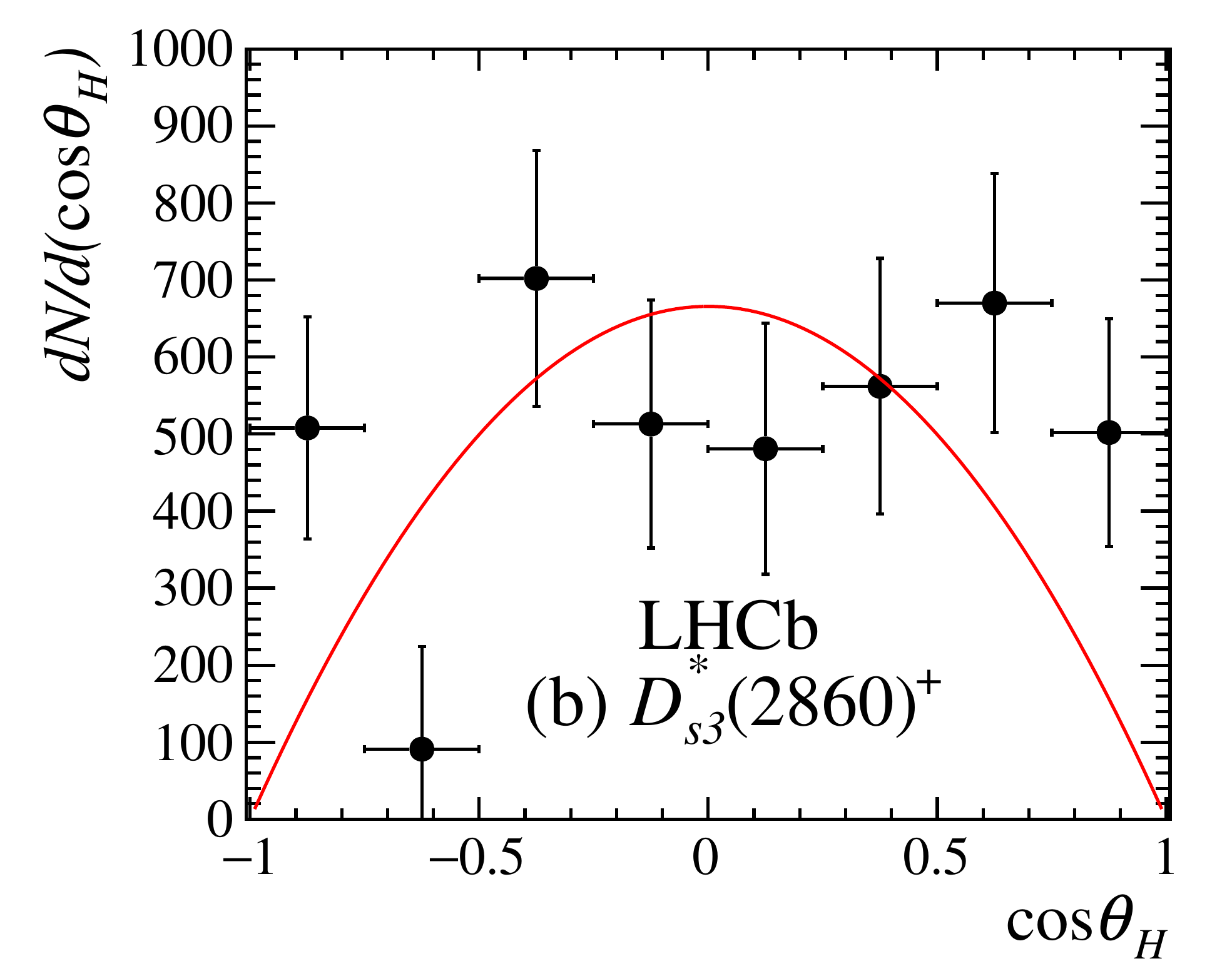}
    \includegraphics[width=0.45\linewidth]{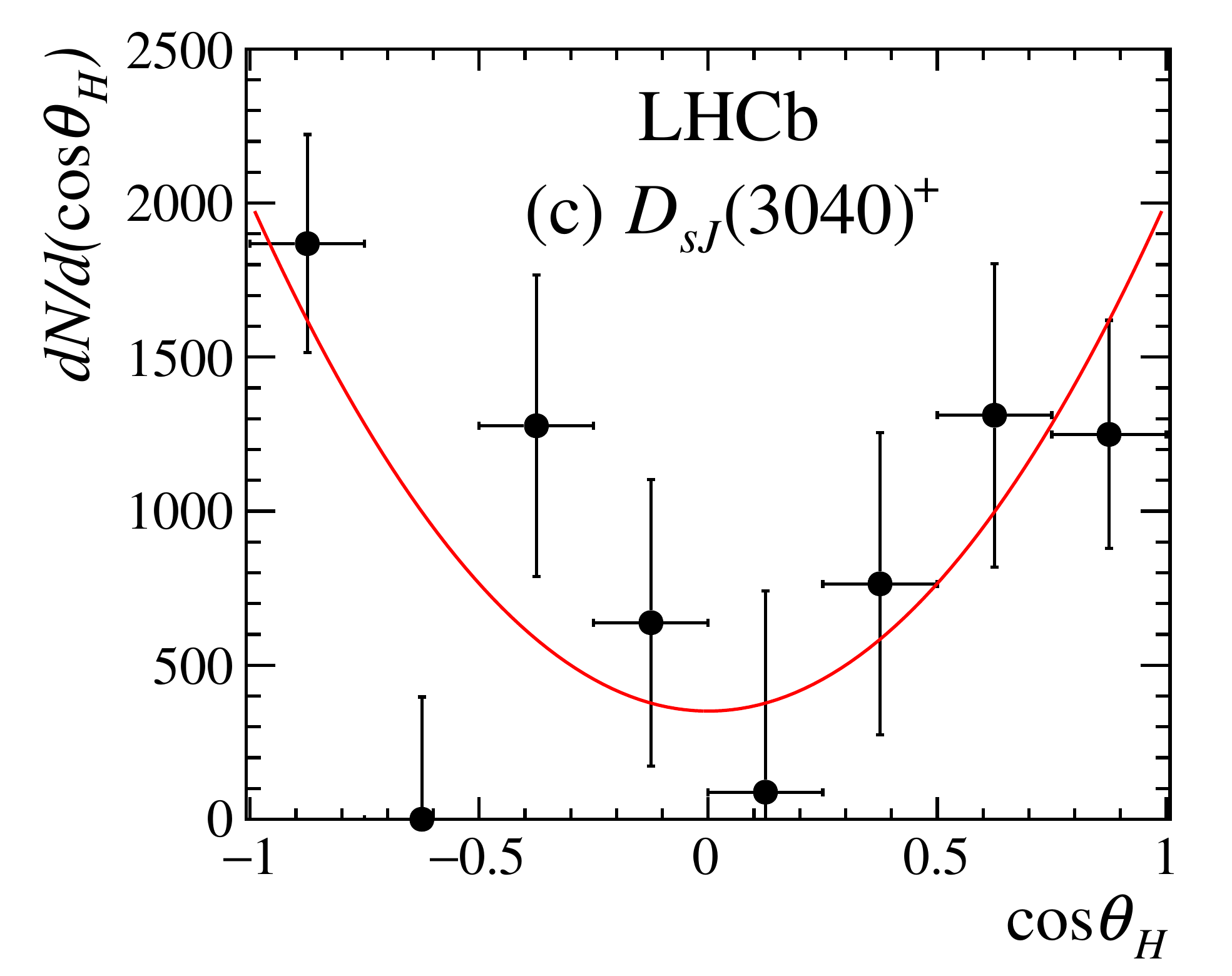}
    \vspace*{-0.5cm}
  \end{center}
  \caption{
      \small Distributions of the measured signal yields for (a) \Dsa, (b) \Dsjb and (c) \Dsc as a function of the 
helicity angle \cthetah. The distributions are fitted with  NP (a,b) and UP (c) functions.
    }
  \label{fig:fig9}
\end{figure}

The $\Dstarp \KS$ data, with \Dzkpi, are then divided into eight equally spaced bins in $\cthetah$. The mass spectra are
fitted (for $m(\Dstarp \KS)<3400\,\mev$) with the model described in Sec.~\ref{sec:mass} with fixed resonance parameters, to obtain the yields as functions of $\cthetah$ for each resonance. The resulting distributions are shown in Fig.~\ref{fig:fig9} and details of the fit results are given in Table~\ref{tab:tab6}.

We observe that the \Dsa state is reasonably well described by the expected NP function ($\chi^2/{\rm ndf}=11.4/7$ with p-value 12.2\%). The fit to the \Dsb angular distribution has a slightly lower p-value (6.3\%). Reference~\cite{Colangelo:2012xi} suggests the possibility of the presence of UP state contributions in this mass range, which cannot be excluded in this fit: there is evidence for the presence of a small signal in the $2860\,\mev$ mass region for the UP sample shown in Fig.~\ref{fig:fig4}(b).
The consistency with the NP assignment confirms the presence of the decay $\Dsb \to \Dstarp \KS$. We also show in Fig.~\ref{fig:fig9}(c) the $\cthetah$ distribution for the enhancement at the \Dsc position and find it consistent with a UP assignment.

\vspace{0.5cm}

\section{Summary}

 A study of the resonant structures in the \dstarks and \dstarzk systems is performed using $pp$ collision data, collected at centre-of-mass energies of 7 and 8~\tev with the \lhcb detector.
   For the  \dstarks final state, the decay chains \dstartodpi with \Dzkpi and \Dzkpia are used, with an integrated luminosity of $3.0\,\invfb$. For \dstarzk, the decay chain \dstarztodpi, \Dzkpi is used, with an
  integrated luminosity of $2.0\,\invfb$.

A prominent \Dsone  resonance is observed in both \dstarks and \dstarzk final states.
Resonances \Dsa and \Dsb are also observed and their parameters are measured to be
  \begin{align*}
    m(\Dsa) &= 2732.3 \pm 4.3\stat \pm 5.8\syst\,\mev, \nonumber \\
    \Gamma(\Dsa) &= 136 \pm 19\stat \pm 24\syst\,\mev, \nonumber 
    \end{align*}
    and
  \begin{align*}
    m(\Dsjb) &= 2867.1 \pm 4.3\stat \pm 1.9\syst\,\mev, \nonumber\\
    \Gamma(\Dsjb) &= 50 \pm 11\stat \pm 13\syst\,\mev \nonumber.
    \end{align*}

Study of the angular distributions supports natural parity assignments for both resonances, although  the presence of
an additional unnatural parity contribution in the $2860\,\mev$ mass range cannot be excluded.
The data are not sensitive to the presence of an additional \Dsjonep resonance.

The \Dstwo decay to \dstarks is also observed for the first time, at a significance of $6.9\,\sigma$, with a branching fraction relative to the $\Dp \KS$ decay mode of
\begin{equation}
\frac{\calB(\Dstwo \to \Dstarp \KS)}{\calB(\Dstwo \to \Dp \KS)} = 0.044 \pm 0.005\stat \pm 0.011\syst.
\end{equation}
 This measurement is in agreement with expectations from recent calculations of the charm and charm-strange mesons spectra~\cite{Godfrey:2015dva} which predict a value of 0.058 for this ratio.
A spin-parity analysis of the decay $\Dstwo \to \Dstarp \KS$ supports the natural parity assignment.
The data also show weak evidence for further structure in the region around $3040\,\mev$ consistent with contributions from unnatural parity states. 
\section*{Acknowledgements}

\noindent We express our gratitude to our colleagues in the CERN
accelerator departments for the excellent performance of the LHC. We
thank the technical and administrative staff at the LHCb
institutes. We acknowledge support from CERN and from the national
agencies: CAPES, CNPq, FAPERJ and FINEP (Brazil); NSFC (China);
CNRS/IN2P3 (France); BMBF, DFG and MPG (Germany); INFN (Italy); 
FOM and NWO (The Netherlands); MNiSW and NCN (Poland); MEN/IFA (Romania); 
MinES and FANO (Russia); MinECo (Spain); SNSF and SER (Switzerland); 
NASU (Ukraine); STFC (United Kingdom); NSF (USA).
We acknowledge the computing resources that are provided by CERN, IN2P3 (France), KIT and DESY (Germany), INFN (Italy), SURF (The Netherlands), PIC (Spain), GridPP (United Kingdom), RRCKI and Yandex LLC (Russia), CSCS (Switzerland), IFIN-HH (Romania), CBPF (Brazil), PL-GRID (Poland) and OSC (USA). We are indebted to the communities behind the multiple open 
source software packages on which we depend.
Individual groups or members have received support from AvH Foundation (Germany),
EPLANET, Marie Sk\l{}odowska-Curie Actions and ERC (European Union), 
Conseil G\'{e}n\'{e}ral de Haute-Savoie, Labex ENIGMASS and OCEVU, 
R\'{e}gion Auvergne (France), RFBR and Yandex LLC (Russia), GVA, XuntaGal and GENCAT (Spain), The Royal Society, Royal Commission for the Exhibition of 1851 and the Leverhulme Trust (United Kingdom).

\addcontentsline{toc}{section}{References}
\bibliographystyle{LHCb}

\bibliography{main}

\clearpage

\centerline{\large\bf LHCb collaboration}
\begin{flushleft}
\small
R.~Aaij$^{39}$, 
C.~Abell\'{a}n~Beteta$^{41}$, 
B.~Adeva$^{38}$, 
M.~Adinolfi$^{47}$, 
A.~Affolder$^{53}$, 
Z.~Ajaltouni$^{5}$, 
S.~Akar$^{6}$, 
J.~Albrecht$^{10}$, 
F.~Alessio$^{39}$, 
M.~Alexander$^{52}$
% ...

{\footnotesize \it
% ...
$ ^{5}$Clermont Universit\'{e}, Universit\'{e} Blaise Pascal, CNRS/IN2P3, LPC, Clermont-Ferrand, France\\
$ ^{6}$CPPM, Aix-Marseille Universit\'{e}, CNRS/IN2P3, Marseille, France\\
% ...
$ ^{10}$Fakult\"{a}t Physik, Technische Universit\"{a}t Dortmund, Dortmund, Germany\\
% ...
$ ^{38}$Universidad de Santiago de Compostela, Santiago de Compostela, Spain\\
$ ^{39}$European Organization for Nuclear Research (CERN), Geneva, Switzerland\\
% ...
$ ^{41}$Physik-Institut, Universit\"{a}t Z\"{u}rich, Z\"{u}rich, Switzerland\\
% ...
$ ^{47}$H.H. Wills Physics Laboratory, University of Bristol, Bristol, United Kingdom\\
% ...
$ ^{52}$School of Physics and Astronomy, University of Glasgow, Glasgow, United Kingdom\\
$ ^{53}$Oliver Lodge Laboratory, University of Liverpool, Liverpool, United Kingdom\\
% ...
}
\end{flushleft}

\end{document}